\documentclass[12pt]{emulateapj}

\newcommand{\Msun}{$M_{\sun}$}

\slugcomment{Accepted for Publication in ApJ}
\shorttitle{COSMOS Bars}
\shortauthors{Sheth et al.}

\begin{document}

\title{Evolution of the Bar Fraction in COSMOS: Quantifying the Assembly of the Hubble Sequence}

\author{Kartik Sheth \altaffilmark{1,2}, Debra Meloy Elmegreen
  \altaffilmark{3}, Bruce G. Elmegreen\altaffilmark{4}, Peter Capak
  \altaffilmark{2}, Roberto G. Abraham \altaffilmark{5}, E.
  Athanassoula \altaffilmark{6}, Richard S. Ellis \altaffilmark{2},
  Bahram Mobasher \altaffilmark{7}, Mara Salvato\altaffilmark{2}, Eva
  Schinnerer\altaffilmark{8}, Nicholas Z. Scoville\altaffilmark{2},
  Lori Spalsbury \altaffilmark{2}, Linda Strubbe\altaffilmark{9}, 
  Marcella Carollo\altaffilmark{10}, Michael Rich\altaffilmark{11},
Andrew A.  West\altaffilmark{9}}

\altaffiltext{1}{Spitzer Science Center, California Institute of Technology, Pasadena, CA 91125}
\altaffiltext{2}{California Institute of Technology, MC 105-24, 1200 East California Boulevard, Pasadena, CA 91125}
\altaffiltext{3}{Department of Physics and Astronomy, Vassar College, 124 Raymond Avenue, Poughkeepsie, NY 12604}
\altaffiltext{4}{IBM T. J. Watson Center, P.O. Box 218, Yorktown Heights, NY 10598}
\altaffiltext{5}{Department of Astronomy \& Astrophysics, University of Toronto, 50 St. George St, Room 1403, Toronto, ON M5S 3H4, Canada}
\altaffiltext{6}{LAM, OAMP, 2 place Le Verrier, 13248 Marseille cedex 04, France}
\altaffiltext{7}{Space Telescope Science Institute, 3700 San Martin Drive, Baltimore, MD 21218}
\altaffiltext{8}{Max Planck Institut f\"ur Astronomie, K\"onigstuhl 17, Heidelberg, D-69117, Germany}
\altaffiltext{9}{Astronomy Department, University of California, 601 Campbell Hall, Berkeley, CA 94720-3411}
\altaffiltext{10}{Department of Physics, ETH Zurich, CH-8093 Zurich, Switzerland}
\altaffiltext{11}{Department of Physics and Astronomy, University of California, Los Angeles, CA 90095}

\begin{abstract}

  We have analyzed the redshift-dependent fraction of galactic bars
  over 0.2$<$z$<$0.84 in 2,157 luminous face-on spiral galaxies from
  the COSMOS 2-square degree field.  Our sample is an order of
  magnitude larger than that used in any previous investigation, and
  is based on substantially deeper imaging data than that available
  from earlier wide-area studies of high-redshift galaxy morphology.
  We find that the fraction of barred spirals declines rapidly with
  redshift.  Whereas in the local Universe about 65\% of luminous
  spiral galaxies contain bars (SB$+$SAB), at $z\sim0.84$ this
  fraction drops to about 20\%. Over this redshift range the fraction
  of {\em strong} (SB) bars drops from about 30\% to under 10\%.  It
  is clear that when the Universe was half its present age, the census
  of galaxies on the Hubble sequence was fundamentally different from
  that of the present day.  A major clue to understanding this
  phenomenon has also emerged from our analysis, which shows that the
  bar fraction in spiral galaxies is a strong function of stellar
  mass, integrated color and bulge prominence. The bar fraction in
  very massive, luminous spirals is about constant out to z$\sim$0.84
  whereas for the low mass, blue spirals it declines significantly
  with redshift beyond z=0.3.  There is also a slight preference for
  bars in bulge dominated systems at high redshifts which may be an
  important clue towards the co-evolution of bars, bulges and black
  holes.  Our results thus have important ramifications for the
  processes responsible for galactic downsizing, suggesting that
  massive galaxies matured early in a dynamical sense, and not just as
  a result of the regulation of their star formation rate.

\end{abstract}


\section{Introduction}\label{intro}

How, when and at what rate did the Hubble sequence form?  This
question is central to the field of galaxy formation and evolution. We
examine it by measuring the evolution of the bar fraction with
redshift using the 2-square degree Cosmic Evolution Survey (COSMOS).
In nearly all simulations, the formation timescale for a bar is rapid
{\sl once the necessary conditions (a massive, dynamically cold and
  rotationally-supported disk)} are met.  Therefore the redshift
evolution of the bar fraction is a fundamental probe of the
evolutionary history of disk galaxies.

The bar fraction is defined simply as:
\begin{equation}
f_{\rm bar} = {{\rm number \ of\  barred\  spirals}\over {\rm number\  of\  all\ spirals}}.
\end{equation}
In the local Universe the value of $f_{bar}$ is quite well
established.  When only strongly barred\footnote{Bars that are highly
  elliptical and have rectangular isophotes are classified as strongly
  barred (SB) galaxies whereas those with more oval shapes are
  classified as SAB or ovally distorted galaxies} galaxies (SB) are
counted, the RSA, RC3 and UGC \citep{sandage87,devau91,nilson73} all
give values of $f_{bar}=0.25-0.3$. When ovally distorted (SAB) are
also counted the situation becomes a little less clear-cut, because,
unlike the RC3, the UGC and RSA do not attempt to carefully compile an
inventory of such galaxies. If ovally distorted systems in the RC3 are
included in the computation of $f_{bar}$ then the local bar fraction
rises to $f_{bar}\sim0.6$. This result is in good agreement with
recent infrared studies which have measured the local bar fraction to
be $\sim$ 0.65 \citep{eskridge00, whyte02, menendez07, marinova07}.
In the infrared, a majority of the SAB galaxies are classified as
strongly barred SB systems \citep{eskridge00}.  As noted by
\citet{eskridge00} and \citet{menendez07}, the overall bar fraction is
the same in the infrared and the optical (although there is a small
number of cases where bars are unveiled at infrared wavelengths).
This is not surprising, because bars are primarily stellar structures
whose visibility only declines sharply at ultraviolet wavelengths,
short wards of the Balmer break (see also \S \ref{bandshifting}).  We
conclude that the consensus value of the local barred fraction is
$f_{bar}\sim0.3$ for strongly barred systems, and $f_{bar}\sim0.65$
for all barred galaxies, and that these values are so well-known that
they have not changed significantly in over four decades.

In sharp contrast with the rapid and stable consensus reached on the
local bar fraction, attempts to measure the bar fraction at high
redshift have proven difficult.  The earliest analyses of the bar
fraction in the Hubble Deep Fields (HDFs) found a dramatic paucity of
barred spirals at z$>$0.5 \citep{abraham96, vanden96, abraham99}.
These authors concluded that at lookback times greater than 5 Gyr
disks were either dark matter dominated or dynamically too hot
(perhaps due to the increased merging activity) to host bars.
However, the small volume probed by the HDFs (only thirty bright,
face-on spiral galaxies between $0<z<1$) led to concerns that the bar
fraction at high redshift may not be adequately measured.
\citet{sheth03} investigated whether a significant number of bars
could have been missed, as suggested by \citet{bunker99} using the
H-band NICMOS HDF.  \citet{sheth03} found four bars and two candidate
bars out of 95 galaxies at z$>$0.7.  Overall, the fraction of barred
spirals in the NICMOS HDF remained extremely low, as in the optical
HDF studies.  But \citet{sheth03} noted that their study was limited
by the coarse NICMOS resolution (0$\farcs$15) such that only the
largest (and rarest) bars could be identified (bars with semi-major
axis $>$5 kpc).  When the fraction of these large bars at z$>$0.7 was
compared to local samples, there was no compelling evidence for a
decline in barred spirals, but likewise the NICMOS data did not unveil
any new bars at low redshifts; all except one of the four bars in the
\citet{sheth03} study are at z$>$0.9, where k-correction effects are
important (\S \ref{bandshifting}).

A major advance in spatial resolution was possible with the Advanced
Camera for Surveys (ACS) whose 0$\farcs$05 pixels are able to resolve
all but the smallest (nuclear, $<$2 kpc diameter) bars at all
redshifts.  Using ACS data, two studies \citep{elm04,jogee04} found
that contrary to the previous HDF results, the bar fraction is
constant at 30\% over the last 8 Gyr (since $z=1.2$).  The sample
sizes, however remained modest in these studies (186 in
\citealt{elm04}, and 258 in \citealt{jogee04}). 

In this paper we examine in detail the redshift evolution of the bar
fraction using the unparalleled wide and deep 2-square degree COSMOS
data set.  The plan for the paper is as follows: in Section 2 we describe our
sample selection procedure. The classification methodology we have
adopted is described in Section 3. Our main results are presented in
Section 4, before being discussed in Section 5. Our conclusions are
summarized in Section 6. An Appendix to this paper provides a detailed
analysis of possible selection effects at high redshift and a
discussion of our local calibration sample of 139 galaxies from the
Sloan Digital Sky Survey (SDSS) Data Release 4 \citep{adelman06}.
Throughout this paper we adopt a flat $\Lambda$-dominated cosmology
with $H_0$=70 km\ s$^{-1}$\ Mpc$^{-1}$, $\Omega_M=0.3$, and
$\Omega_\Lambda=0.7$.

\section{Sample Selection}\label{selection}

An overview of the COSMOS program is given in \citet{scoville07a} and
details of the HST observations are described in
\citet{scoville07b}. The COSMOS observations are undertaken in the
F814W filter (`I-band') and reach a depth of I$_{AB} >$ 27
(10$\sigma$).  The photometric catalog and redshift measurements used
in this paper are given in \citet{mobasher07} and \citet{capak07}.

The most important and difficult step in studying the evolution of
galactic structures is choosing comparable samples at different
lookback times.  For nearby galaxies, multi-waveband data with
sufficient spatial resolution are available.  Therefore galaxy
properties (e.g., Hubble type, inclination, distance) are accurately
known for nearby spirals.  The underlying disk is also well-imaged and
multiple techniques for identifying a bar may be employed.  However,
for high redshift galaxies, the situation is more complicated.  Here
we summarize the steps we have taken to overcome these problems.

We choose all galaxies brighter than L$_V^{*}$ with an empirically
determined luminosity evolution of 1 magnitude from \citet{capak03},
such that M$_V^{*}$ = -21.7 at z=0.9 \citep{capak03}.  This criterion
is specifically targeted for choosing galaxies from the same portion
of the galaxy luminosity distribution at all redshifts.  As we shall
see later, a no-evolution luminosity model would have only steepened
the overall decline in the bar fraction, further strengthening the
results presented in this paper.  For this luminosity criterion, at a
redshift of z=0.9, the sample is complete for all galaxies with a
half-light radius smaller than 10 kpc (see Figure 6,
\citealt{scoville07b}).  The number of galaxies with a half light
radius larger than 10 kpc is extremely low (e.g., Figure 10 in
\citealt{sargent07}, or Figure 10 in \citealt{barden05}) and thus our
sample is essentially complete.

We eliminate all elliptical and lenticular galaxies based on the
galaxy's spectral energy distribution (SED) type T$_{phot}$, choosing
all galaxies with T$_{phot} >$ 2.  T$_{phot}$ is the best fit spectral
template ordered by the 4000\AA\ break strength (see
\citealt{mobasher07}).  The T$_{phot}$ sequence can be thought of as a
photometric Hubble type going from the reddest early type
(T$_{phot}$=1) galaxies to bluest late type (T$_{phot}$=6) galaxies.
Types 1 through 4 are defined by the templates from \citet{coleman80} and
correspond to Elliptical, Sbc, Scd, and Irregular Hubble types
respectively.  Types 5 and 6 correspond to \citet{kinney96} type SB3
and SB2 respectively which are local star-burst galaxies with little
or no extinction.  Typical uncertainties in T$_{phot}$ are $\pm$0.2.

The T$_{phot}$ values are robust descriptors of galaxies to z$\sim$1.2
as confirmed by a comparison of the photometric and spectroscopic
redshifts for over eight hundred galaxies \citep{mobasher07}.
Comparison of galaxy types based on morphological parameters such as
Gini and asymmetry and the T$_{phot}$ type shows that T$_{phot} > 2$
selects all spirals to z$\sim$1.2 \citep{capak07,ilbert07}.  With
increasing redshift there is an increase in the population of blue
ellipticals and blue-bulge dominated spirals; however these galaxies
are only significant at the faint end of the luminosity function
(e.g., \citealt{ilbert06,capak07,ilbert07}).  For instance, at M$_B <
$-20, the blue-bulge dominated spiral population is less than 1\% of
the total disk population \citep{ilbert06}.  At z$\sim$0.8, amongst
elliptical galaxies (identified by the Gini and asymmetry parameters),
the contribution of blue ellipticals to the total volume density is
significant at the very low mass end ($\sim$10$^{9}$ \Msun).  For the
high luminosity (massive) galaxies, which are studied in this sample,
the contribution from blue ellipticals is no more than a few percent.
We do find a small fraction of compact systems with blue colors which
we identify and reject from our analysis.

We impose a redshift cutoff of $z = 0.835$ so that the F814W filter
does not probe bluer than rest-frame g-band. The rationale for probing
no bluer than rest-frame g-band is described in detail in
\S\ref{bandshifting} of the Appendix, but we also note here that at
this redshift the angular diameter of a 0.05$\arcsec$ ACS pixel
subtends a physical scale of 0.4 kpc.  At this resolution we expect to
detect easily the entire population of bars in nearby spiral galaxies
(see Figure 3 of \citealt{sheth03}, or Figure 7 of
\citealt{menendez07}).  We note that all our galaxies subtend at least
10 pixels.  The smallest petrosian radius in our sample is 5.2 pixels.

Imposing these cuts reduces our sample to 3886 spirals. However,
galactic structures like bars are difficult to identify and quantify
in edge-on galaxies.  Therefore, we further eliminate all galaxies
with inclinations $i >$ 65$^{\circ}$, the same limiting inclination
used in studies of the bar fraction in nearby spirals (e.g.,
\citealt{menendez07}).  Inclination values for our sample are
determined from the axial ratio of the galactic disk, which is
identified using a two dimensional decomposition with a bulge and
exponential component for each galaxy using GALFIT \citep{peng02}.
The inclination cut eliminates an additional 986 objects, leaving 2900
galaxies.

Finally, we discard all galaxies that were obviously merging, or too
irregular or peculiar to be fit with ellipses.  This eliminates 743
galaxies, leaving a final sample of 2,157 spiral galaxies which were
then classified as barred or unbarred.  The importance of eliminating
peculiar objects from our analysis is investigated further in
\S\ref{pec} of the Appendix, but we note here that including these
objects would not have changed any of our conclusions.

All galaxies are detected in at least thirteen, sometimes all sixteen
available photometric bands.  The galaxy luminosities, masses and
colors are measured from the photometric redshift code
\citep{mobasher07} which simultaneously solves for observed galaxy
flux, redshift, galaxy type, and extinction intrinsic to the galaxy.
The galaxy type combined with the flux-normalization yields rest
frame, extinction corrected, luminosities and colors for each object.
The rest frame color and luminosity is then converted into stellar
mass using the \citet{bell05} relation.  Errors in the mass
estimate are within a factor of 3 due to systematics.  With our deep
multi-wavelength photometry these parameters are robust and
non-degenerate at z$<$1.2. Over this redshift range we have at
least two data points red-ward of the 4000 \AA\ break ($z^+$ and $K_s$
at z=1.2) and two points blue-ward ($u^*$ and $B_J$ at z=0.3), which
allows us to break the 4000\AA\ break strength-extinction degeneracy.
More detailed SED fits using spectroscopic redshifts yield similar
results for the galaxy luminosity, mass and color \citep{mobasher07,
  ilbert07}.

\section{Bar Classification Methodology}\label{methods}

We identified bars using two methods and the results were
cross-checked for consistency.  We use the same methods for the local
SDSS calibration sample (see \ref{bandshifting}) and the COSMOS data
to reduce biases that can be induced by the use of different
classification methods.

Our first method was to use the ellipse fitting technique that has
been used widely by multiple studies of both nearby and high redshift
galaxies (e.g., \citealt{regan97, sheth00, knapen00, sheth02, laine02,
  sheth03, jogee04, menendez07, marinova07}).  For a detailed
discussion of the ellipse fitting method and the classification scheme
we refer the reader to \citet{menendez07}.  Briefly, bars are
identified from a dual signature of the ellipticity and position angle
profiles.  We require that the ellipticity increases monotonically,
exceeds 0.2, and then drops by at least $\delta \epsilon >$0.1.  The
position angle profile should show a relatively constant position
angle over the bar region and change by at least 10$^o$ after the bar
as the isophotes enter the disk.  In general the fitting procedure is
extremely robust and fits ellipses to the 1$\sigma$ noise level in the
images -- in the case of COSMOS this means we are able to fit galaxies
to the outermost disk isophotes (at $\mu$=24.5 mag arcsec$^{-2}$, see
Figure 9) in the highest redshift bins.  In a small number of cases,
the ellipse fitting method can miss an existing bar (see Figure 3 in
\citealt{menendez07}).  This usually happens when the position angle
of the disk and bar are aligned or when the ellipticity drop is
moderated by open spiral arms.

In addition to the ellipticity and position angle profiles, we
classified the bars further into strong bars and intermediate bars by
visually examining the isophote shapes.  The strong bars are those
with an ellipticity greater than 0.4 to be consistent with previous
work in this field (e.g., \citealt{jogee04}).  However we note that
the definition is arbitrary because there is a continuum of bar
strengths \citep{block02, whyte02, buta04, buta05, menendez07}.  We
refer to this first method as the `ElPa' classification method for the
remainder of the paper.

Our second method for classifying galaxies was visual identification
of galaxies into categories of SA (unbarred), SAB, and SB by one of the
authors (DME).  Not all galaxies were classified into these three neat
categories.  DME also classified galaxies as edge-on, clump-cluster
galaxy, or compact (spheroidal) galaxy following her work on the Ultra
Deep Field \citep{elm05a,elm05b,elm05c,elm06, elm07}.  Galaxies that
were visually classified as edge-on, clump-cluster, or compact are not
included in the visual classification results.  The total number of
galaxies classified visually into barred and unbarred spirals is
1,705.

The classification by DME was cross-checked for five hundred galaxies
by another author (KS), and also cross-checked against the ellipse
fitting method.  The cross checks between the two methods finds
agreement for 85\% of the sample.  In general we found that we
classified more galaxies as barred by eye than using the ellipse
fitting profile.  This is as expected because there are particular
morphologies where the ellipticity and position angle signature can be
masked by the relative orientation of the bar and disk, and the pitch
angle of the spiral arms.  As noted earlier, a detailed discussion of
such cases can be found in \citet{menendez07}.  Only in 5\% of the
cases is there a gross mismatch where one method differed from the
other by two classes, i.e. bar in one and spiral in the other.  This
generally occurred for very faint or small nuclei where the ellipse
fitting method has problems fitting the isophotes.  In the remaining
10\% of the cases, the methods agreed to within one class,
i.e. intermediate bar in one, and a spiral or strong bar in the other.
The cross checks were internally consistent at all redshifts.

\section{Results}

\subsection{The Declining Bar Fraction}
\begin{figure*}
\epsscale{0.82}
\plotone{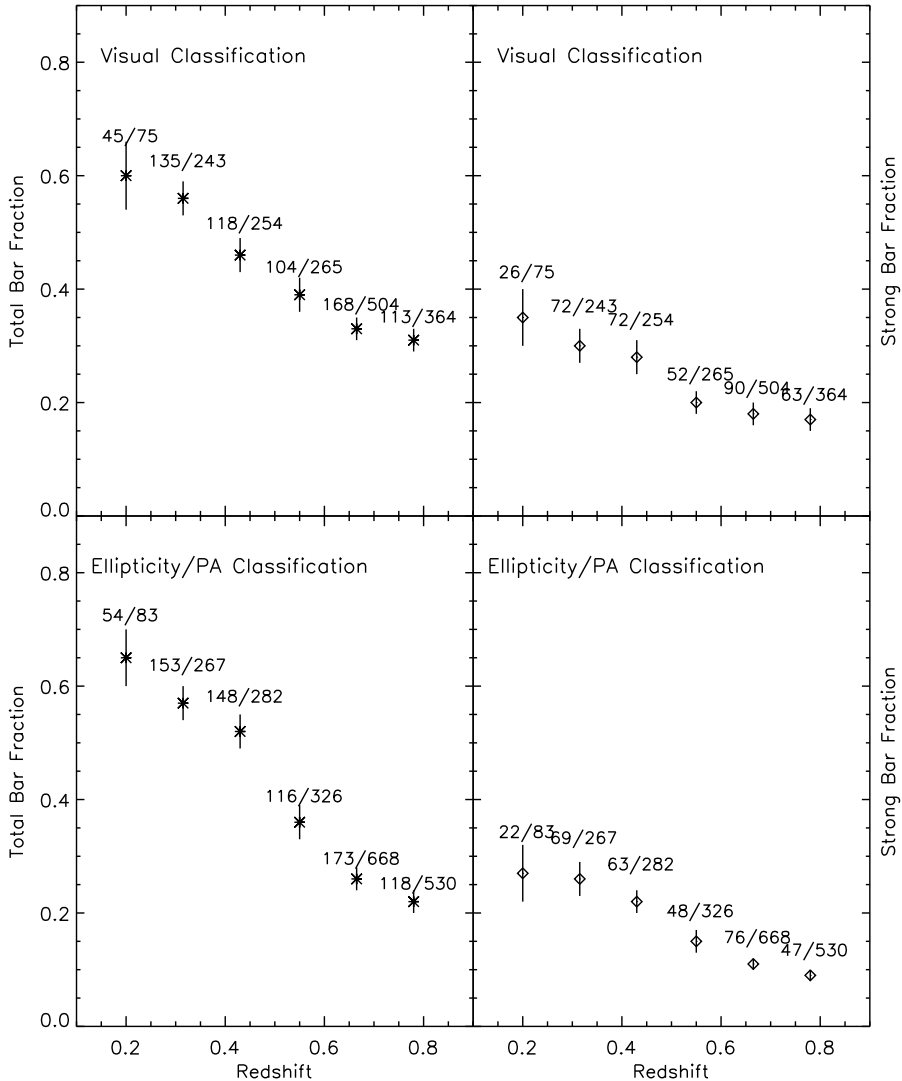}
\caption{Evolution of the bar fraction as a function of redshift in
  equal bins from z=0.0 to z=0.84, out to a look back time of 7
  Gyr. The bar fraction drops from 65\% in the local Universe to about
  20\% at z$\sim$0.84. The fraction of {\em strong} bars (SB) drops
  from about 30\% to under 10\%.  The top row shows the results from
  the visual classification and the bottom row shows the results based
  on classification using the ellipticity and position angle profiles.
  The left column shows the bar fraction for all galaxies classified
  as bars, whereas the right column shows the same only for the strong
  bars.  The error bars are calculated as $\sqrt (f(1-f)/N_{bin}$,
  where f is the fraction of galaxies, and N is the number of galaxies
  in a given category.  The numbers above each data point show the
  total number of bars (or strong bars) / total number of galaxies in
  the bin. \label{barfrac}}
\end{figure*}

Figure \ref{barfrac} shows the evolution of the bar fraction in COSMOS
as a function of redshift in six equal redshift bins ($\delta$z =
0.117), starting at z=0.14 at a lookback time of 1.8 Gyr. The redshift
bins correspond to lookback times of 1.8--3.0, 3.0--4.1, 4.1--5.0,
5.0--5.8, 5.8--6.5, and 6.5--7.1 Gyr respectively.  The two rows show
the bar fractions measured from the two classification methods
described above (\S \ref{methods}).  The left column shows the total
bar fractions (strong bars $+$ oval bars) and right column shows the
strong-bar fraction (SB).  For each data point we list the number of
bars and the total number of galaxies classified in each bin. The
error bars reflect the statistical uncertainty in the fraction and are
calculated from the expression
$\left(f\left[1-f\right]/N\right)^{1/2}$ for fraction $f$ and number
of galaxies $N$.

We find that the bar fraction (for all galaxy luminosities combined)
has evolved dramatically over the last 7 Gyr.  At z=0.84, the total
bar fraction using the ElPa classification method is f$_{bar}$ = 0.22$\pm$0.02
(0.31$\pm$0.02 for visual classification method), one-third (one-half)
its present-day value.  We see the same trend when we consider only
the strong bar fraction: f$_{SB}$ evolves from 0.27$\pm$0.05
(0.35$\pm$0.05) at z=0.0 to 0.09$\pm$0.01 (0.17$\pm$0.02) at z=0.835.

When combined with the z=0.0 data point from our analysis of a local
SDSS sample (Table 1, Figure \ref{comparestuds}, \& \S
\ref{bandshifting}), we also find that the evolutionary trend is
weaker in the first three bins at $z < 0.4$.  Within the error bars,
the data at these redshifts are consistent with a roughly constant bar
fraction (f$_{bar}$=0.6, f$_{SB}$ = 0.3).  These results are
summarized in Table \ref{tab1}.

\subsection{Bar Fraction as a Function of Galaxy Mass \& Luminosity}
\begin{figure*}
\epsscale{0.75}
\plotone{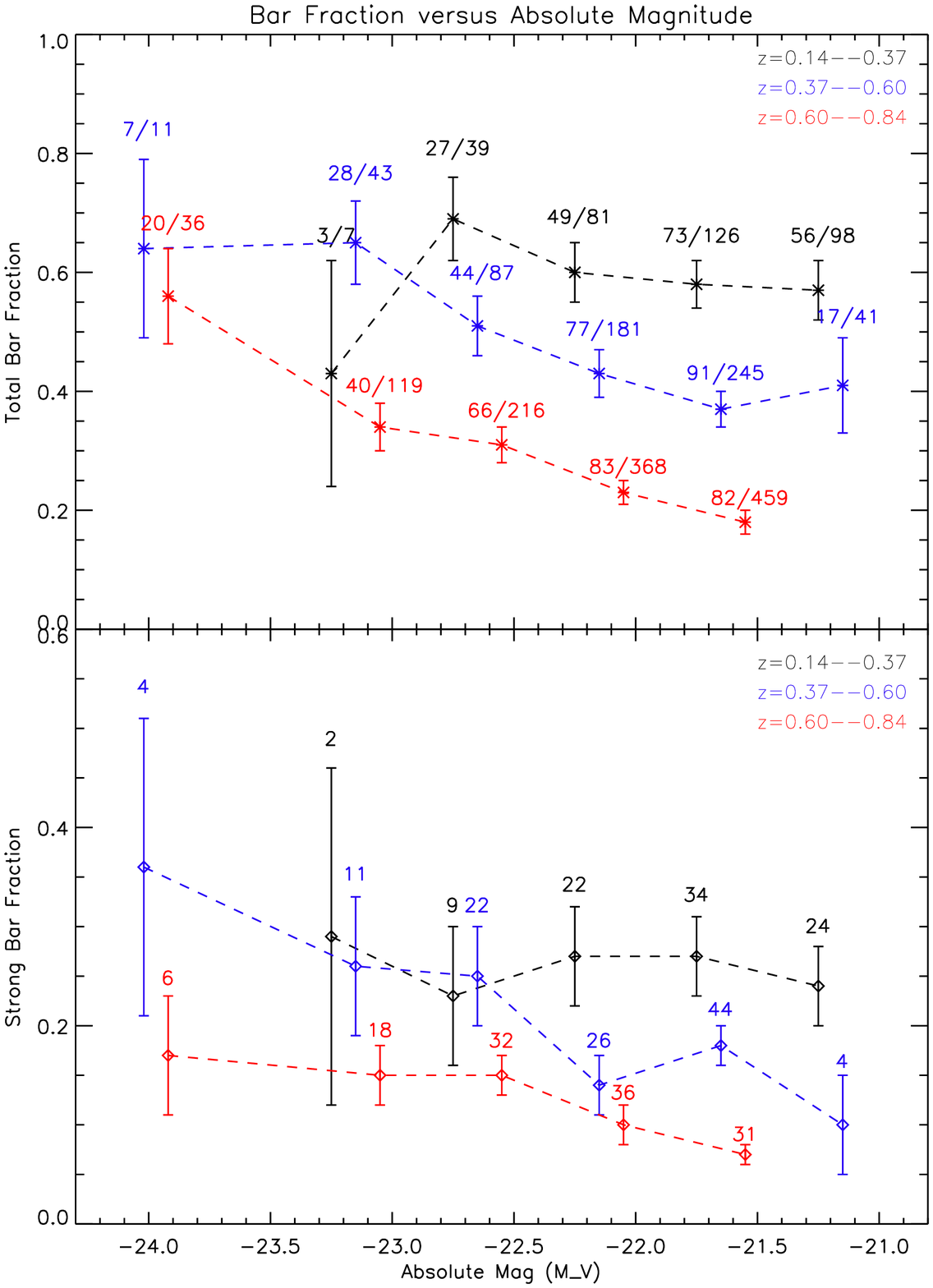}
\caption{Total bar fraction (top panel) and strong bar fraction
  (bottom panel) versus the absolute magnitude (M$_V$) in three
  redshift bins. There is a strong correlation between the bar
  fraction and the galaxy luminosity in the highest redshift bin.  In
  that bin, the most luminous galaxies (M$_V <$ -23.5) have
  f$_{bar}\sim0.5$ whereas the lowest luminosity (M$_V > $ -22.5)
  galaxies have a f$_{bar}\sim0.2$.  With redshift we see a strong
  evolution in the low luminosity sample as they evolve to f$_{bar}
  \sim$0.6 in the lowest redshift bin. A similar trend is seen in the
  bottom panels for the strong bar fraction.  The data points are at
  the mid-points of bins of $\delta$M$_V$=0.5, from -21.0 to -23.5 (a
  data point is skipped if no galaxies are found in a bin) and the
  errors bars are calculated as before. The first bin is from M$_V$ =
  -23.5 to -24.75.  The data points are slightly offset along the
  x-axis for each redshift bin and only the number of strong bars is
  labelled on the bottom panel for clarity. This figure should be
  viewed together with Figure \ref{Barfracmass}. \label{BarfracvsMV}}
\end{figure*}

\begin{figure*}
\epsscale{0.7}
\plotone{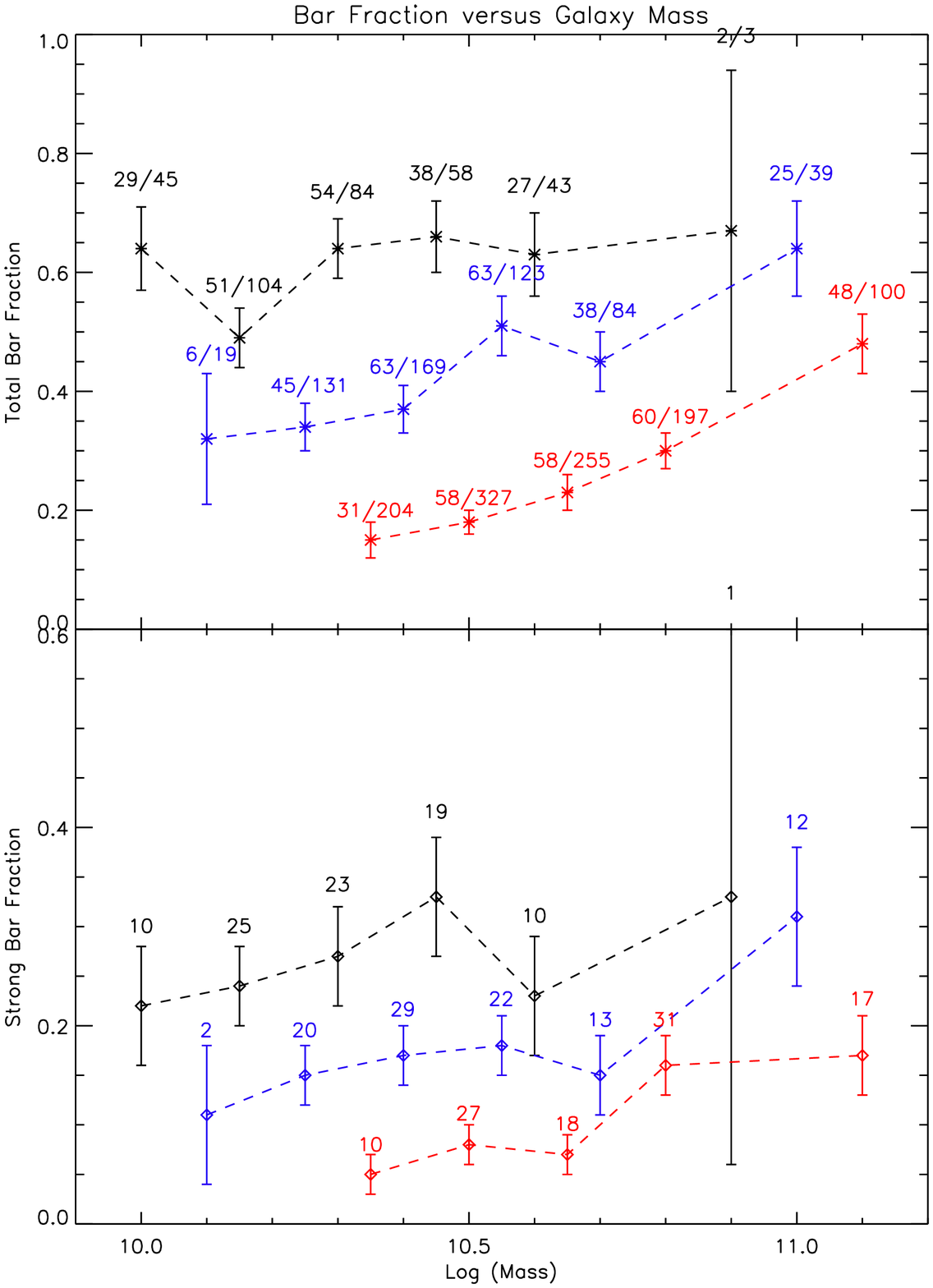}
\caption{Total bar fraction (top panel) and strong bar fraction
  (bottom panel) versus total galaxy stellar mass in three redshift
  bins. As expected from Figure \ref{BarfracvsMV}, there is a strong
  correlation between the bar fraction and the mass for the highest
  redshift bin.  In that bin, galaxies with log M$>$ 10.9 already have
  f$_{bar} \sim0.5$, whereas galaxies with log M $<$ 10.5 have
  f$_{bar} < 0.2$.  The bar fraction for the entire population evolves
  with time with the largest change in the lowest mass bin.  The same
  trend is seen in the bottom panel. The lack of high mass galaxies in
  the lower redshift bins is because, even with 2-square degrees, the
  volume of space observed by COSMOS is small. Luminosity and color
  selection criteria put a limit on the minimum detectable mass - our
  sample is complete for the points shown here.  Each point is at the
  left edge of bins of $\delta$log M=0.15, starting from 10.0 (point
  skipped if no data is found in a bin). Errors bars are calculated as
  before and last bin is from log M = 10.9 to 11.5.  The data points
  are slightly offset along the x-axis for each redshift bin and only
  the number of strong bars is labelled on the bottom panel for
  clarity. The uncertainty in the mass measurement is a factor of
  three.  This figure should be viewed together with Figure
  \ref{BarfracvsMV}. \label{Barfracmass}}
\end{figure*}

Figures \ref{BarfracvsMV} and \ref{Barfracmass} show f$_{bar}$ versus
the absolute luminosity and mass of the disk, respectively, in the
three redshift bins from z=0.14 to z=0.84.  We find that in the
highest redshift bin, galaxies with masses log  M (\Msun)$>$ 10.9 and
luminosities M$_V <$ -23.5 have f$_{bar} \sim$0.5, which is about the
local value.  In contrast, the low mass (log M $<$ 10.5) and
low luminosity (M$_V > $ -22.5) galaxies have f$_{bar} < $ 0.2 at high
redshift.
The same trend is seen for strong bars, f$_{SB}$.  At low redshifts,
the bar fraction is roughly equal for all luminosities. 

This trend is not due to incompleteness in the sample.  We establish
the completeness of our sample by measuring the mass limit based on
our selection criteria.  Since we choose galaxies based on a
luminosity cutoff and galaxy colors, the mass completeness is most
likely to be an issue for the reddest systems at the highest redshift.
For our luminosity cutoff and T$_{phot}$ criteria, our sample is
complete for galaxies with masses greater than 3--4$\times$10$^{10}$
\Msun at z=0.9 for the reddest (T$_{phot}$=2, rest-frame
$\Delta$m$_{g--r} > $ 0.56) galaxies.  Obviously, for the bluest
systems (e.g., T$_{phot}$ = 6), our sample is complete to 0.9--1
$\times$10$^{10}$ \Msun.  These values are calculated from the
\citet{maraston05} and \citet{bell05} models respectively.  Note that
at z=0.6, the mass limit for completeness in the sample is lowered by
another $\sim$25\%.  Our lowest mass data point in the 0.6$<$z$<$0.84
bin (the highest redshift bin) in Figure \ref{Barfracmass} is for
galaxies with masses between 3--4$\times$10$^{10}$\Msun.  It is
therefore free from the possible mass selection bias. The data points
at lower masses and lower redshifts are also computed from a complete
sample of masses for a given bin.  Thus we conclude that the observed
strong correlation between the bar fraction and mass in the highest
redshift bin is a robust result.


The most important result in these figures is that in the highest
redshift bins in this study, a majority of the most massive and
luminous systems are barred. There is little evolution in the bar
fraction with redshift in these systems.  Since bars form in massive,
dynamically cold and rotationally supported galaxies, the high bar
fraction indicates that the most massive systems are already
``mature'' enough to host bars.  This agrees with the analysis of the
evolution of the size function of disk galaxies of several studies
\citep{sargent07, ravindranath04, barden05, sheth07b}, which find that
large disks are already in place by z=1 and little or no evolution in
disk sizes from z$\sim$1 to the present epoch.  Conversely the low bar
fraction in the lower luminosity, lower mass systems indicates that
these systems are either dynamically hot, not rotationally supported
and/or have not accreted sufficient mass to host bars.  Merging
activity, which is more common at higher redshifts, is also likely to
affect the less massive systems more severely and may be responsible
for heating them up more than high mass systems.  Bar formation may be
delayed in these hot disks if they are embedded in a massive dark
matter halo.  Although the exact nature of these disks is not yet
well-known, there is an indication that later type systems may be
dynamically hotter \citep{kassin07}.  We consider these points further
in following sections.

\subsection{Bar Fraction as a Function of Galaxy Color \& Bulge Luminosity}

\begin{figure*}
\epsscale{0.75}
\plotone{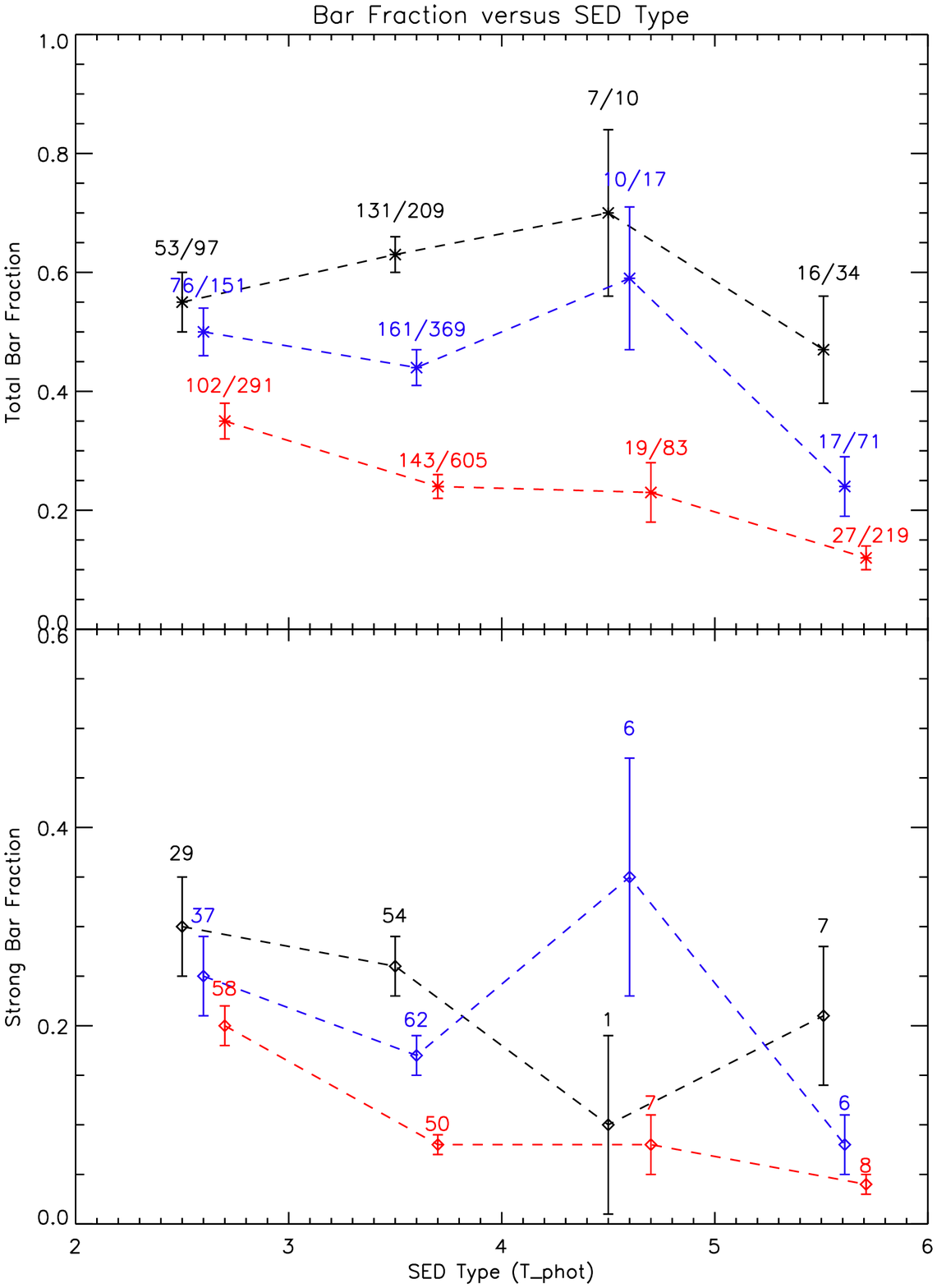}
\caption{Total bar fraction (top panel) and strong bar fraction
  (bottom panel) versus spectral type of a galaxy. In the highest
  redshift bin, the bar fraction decreases monotonically toward later
  spectral types. In contrast, in the lowest redshift bin the bar
  fraction is constant within the error bars across the T$_{phot}$
  sequence.  The data thus show that the majority of the evolution in
  f$_{bar}$ is in the bluer, later SED type systems.  This is
  consistent with the results shown in previous figures for high and
  low mass galaxies, considering the usual correlation between mass
  and T$_{phot}$. As before, the data points are slightly offset along
  the x-axis for each redshift bin and only the number of strong bars
  is labelled on the bottom panel for
  clarity.  \label{Barfracvshubble}}
\end{figure*}

Figure \ref{Barfracvshubble} shows how f$_{bar}$ varies with galaxy
SED type (T$_{phot}$) and redshift. At low redshift, f$_{bar}$ is
independent of T$_{phot}$, and at high redshift, $f_{bar}$ decreases
from early (T$_{phot} < $3) to late types (T$_{phot} > $3).
Similarly, $f_{bar}$ decreases with redshift more strongly for the
late types than the early types. This latter trend is consistent with
the previous result that the bar fraction changes with redshift {\sl
  primarily} for the low mass galaxies, which tend to have late SED
types.

\begin{figure*}
\epsscale{0.75}
\plotone{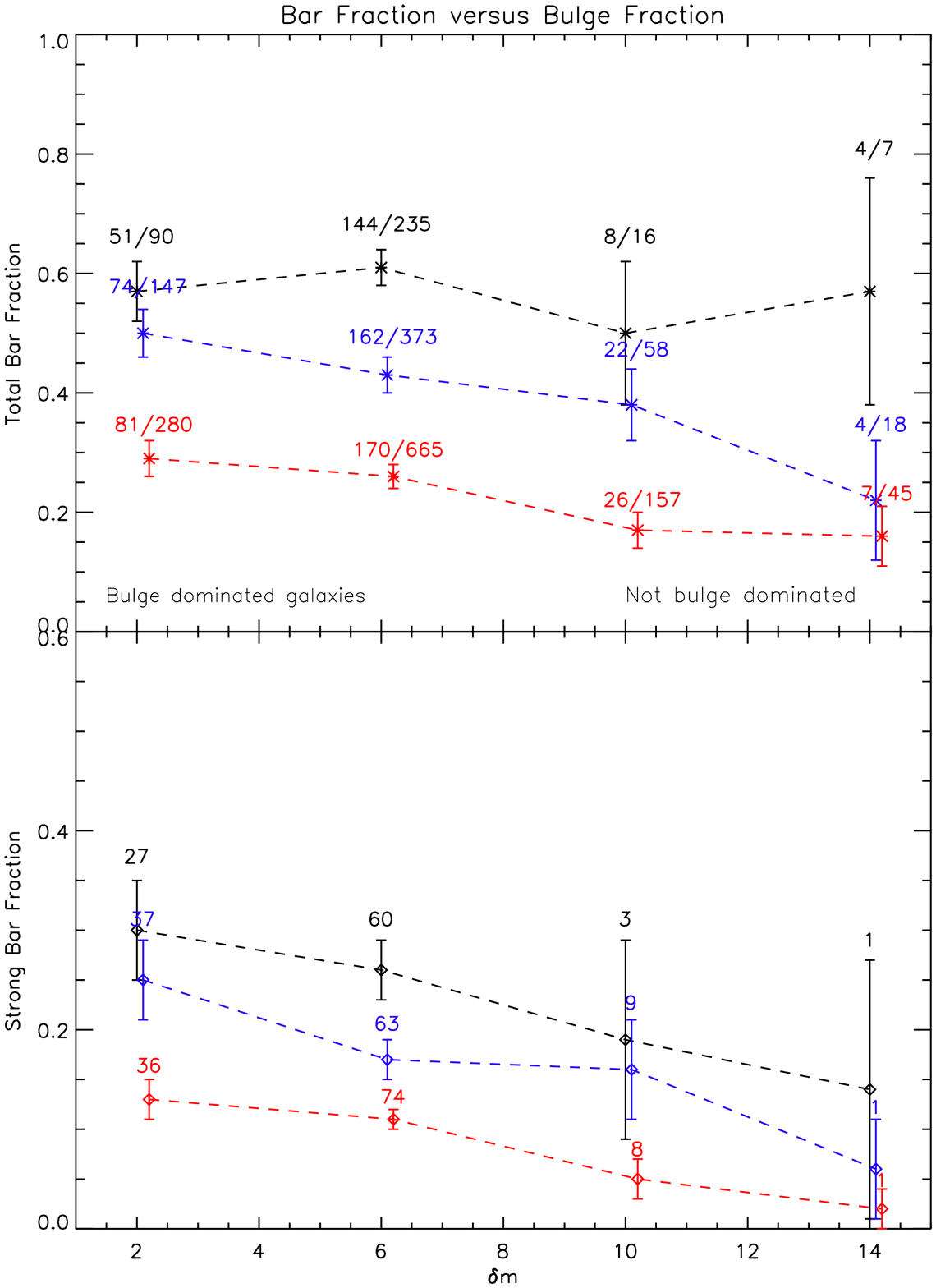}
\caption{Bar fraction as a function of the bulge contribution in
  galaxies in three redshift bins, as before.  We find that in the two
  high redshift bins there is a slight preference for bars to be in
  bulge-dominated systems.  The difference is less pronounced in the
  lowest redshift bin.  The x-axis is calculated by subtracting the
  apparent bulge magnitude from the total galaxy magnitude.  The bulge
  magnitude is calculated by fitting each galaxy with an exponential
  and bulge profile using GALFIT as discussed in the text.  The x-axis
  thus indicates the fractional contribution of the bulge to the total
  luminosity of the galaxy with bulge dominated systems to the left on
  the x-axis.  Since k-corrections are important in two dimensional
  decomposition of galaxies, the results across redshift bins are not
  as robust as those within a given redshift bin.  The data points are
  at the midpoints of each T$_{phot}$ type bin, and as before, only
  the number of strong bars is listed in the bottom
  panel. \label{fbarvsbulge}}
\end{figure*}

Finally we consider how the bar fraction varies as a function of the
bulge light in galaxies.  Figure \ref{fbarvsbulge} shows f$_{bar}$
versus the fraction of bulge luminosity in a galaxy for different
redshift bins.  The bulge magnitude is calculated from fitting each
galaxy with a Sersic $+$ exponential profile using GALFIT
\citep{peng02}.  The x-axis in this figure is the difference between
the bulge magnitude measured from the GALFIT fitting and the total
(disk + bulge) apparent magnitude.  We note that the relative
calibration across redshift bins should be treated with caution
because we are not correcting for k-correction effects that are known
to affect two dimensional decomposition of galaxies. Within a given
redshift bin, however, the bulge contribution measurement should be
robust except for one important caveat.  The fitting algorithm is not
designed to decompose a bar separately.  As a result the bar light is
likely to be split between the exponential and Sersic components.  If
the light profile of a bar is exponential, as it is in later Hubble
type galaxies locally, the majority of that light is likely to be part
of the exponential component.  On the other hand if the bar is
relatively short and not highly elliptical, its light is likely to be
added to the Sersic component.  The detailed decomposition of the
bulge$+$bar$+$disk will require a more sophisticated approach, which is
beyond the scope of this paper.  

Keeping the above caveat in mind, we find that f$_{bar}$ is slightly
higher for galaxies that are ``bulge dominated'' compared to galaxies
which are not bulge dominated in the highest redshift bin.  In the
lowest redshift bin the slight trend disappears and the bar
fraction is roughly constant for all types of bulges, although there
are only a few galaxies that are not bulge dominated.  This correlation
together with the previous correlations (Figures \ref{BarfracvsMV},
\ref{Barfracmass}, \ref{Barfracvshubble}) suggests that the galaxies
that are red, luminous and massive are also bulge dominated, and in
these galaxies the bar fraction does not vary strongly with redshift.
We discuss the implications of the bar-bulge correlation in \S
\ref{bulgeevol}.

\section{Discussion}

\subsection{Comparison with Previous Studies}\label{massevol}

\begin{figure*}
\epsscale{0.8}
\plotone{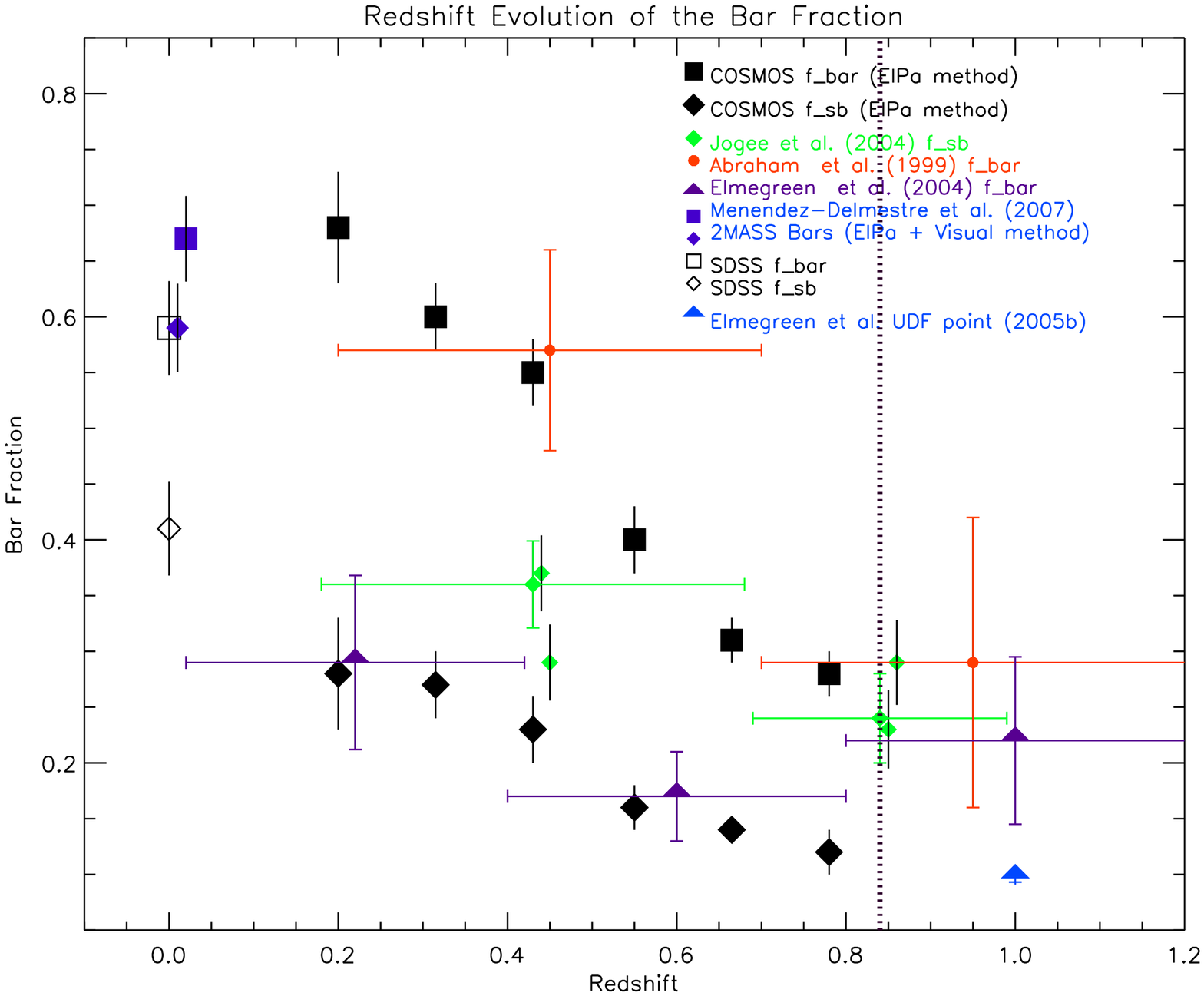}
\caption{Comparison of measurements of our bar fraction with previous
  studies. f$_{bar}$ and f$_{SB}$ measured using the ElPa method for
  the COSMOS data are shown with the black filled squares and diamonds
  respectively. The red data circles (13/30 bars in low z bin, 4/14 in
  the high-z bin) are from \citet{abraham99}; these do not distinguish
  between weak and strong bars. Purple triangles are bars and twists
  from \citet{elm04} (10/34, 15/90, 7/31 bars respectively); we summed
  adjacent bins from their data here and corrected the last bin for
  the incorrect redshifts for four of their galaxies.  The green
  points from \citet{jogee04} are the strong bar fractions from GEMS;
  the three points in the two redshift bins are not independent of
  each other - they are measured for $\sim$110-175 galaxies, chosen in
  different ways from the same sample.  The horizontal bars show the
  redshift range over which these data are averaged.  Also shown are
  data from our analysis of a SDSS control sample (diamond -
  f$_{bar}$, square - f$_{SB}$), and from the 2MASS survey by
  \citet{menendez07} (blue diamond is when both ellipticity and
  position angle signatures are present, and the square also includes
  candidate bars. The vertical dotted line is the limiting redshift
  for our survey. Within the error bars all the data seem to be in
  agreement (although the \citet{jogee04} data point for the lower
  redshift bin is significantly above the other measurements of
  f$_{SB}$).  Contrary to earlier interpretations, it seems that all
  studies show a general decline in the bar fraction with redshift.
  It is only with the COSMOS data that we are able to analyze this
  decline in detail.
\label{comparestuds}}
\end{figure*}

In Figure \ref{comparestuds} we plot all of the data from previous
studies of the evolution of the bar fraction for comparison to the
COSMOS results.  In all cases there is a general decline in the bar
fraction.  However the interpretation of the data has been very
different amongst these studies as we discuss below.

Our basic result of a decline in the bar fraction is consistent with
the earliest HDF studies \citep{abraham96,vanden96,abraham99} shown
with the red data points in Figure \ref{comparestuds}.  These studies
reported a striking decline in the fraction of bars at z$>$0.5.  But
re-analysis of their Figure 4 shows that, in fact, at z$>$0.5, there
are 10 barred spirals out of 29 galaxies, and at z$>$0.8, there are 3
barred spirals out of 11 galaxies, consistent with the COSMOS results
presented here.  Similarly our result is consistent with the very low
bar fraction (5--10\%) measured from the NICMOS HDF \citep{sheth03} at
z$>$0.7.  But it is difficult to compare this bar fraction to the ours
because the NICMOS data can only probe the largest bars.  Of course
the volume probed by the HDF studies was too small to allow evolution
in $f_{bar}$ to be probed with much confidence.

On the other hand, our central result is fundamentally different from
that reported by \citet{jogee04} and \citet{elm04}; these studies have
reported a constant bar fraction with redshift to z$\sim$1.  In the
\citet{jogee04} study, the authors classified 258 galaxies as either
strongly barred or unbarred. Their modest sample size however
prevented them from studying the f$_{bar}$ evolution in detail.  Their
first bin, for example, encompasses our central four redshift bins.
Nevertheless we can compare their measurement of f$_{SB}$ to ours.
Over the same redshift range in the COSMOS data, we measure a strong
bar fraction, f$_{SB}$ = 0.23$\pm$0.01 for the visual classification,
and 0.17$\pm$0.01 for the ElPa classification methods.  In comparison,
\citet{jogee04} reports a f$_{SB} \sim$ 0.3$\pm$0.03\footnote{Errors
  are not reported in the \citet{jogee04} study.  We measure an error
  for their data using the bar fraction and total number of galaxies
  reported, in the same way as we measured for the COSMOS data}.  Even
though the \citet{jogee04} study probes fainter galaxies (M$_V <
$-19.3), which should have resulted in a lower bar fraction, their
f$_{SB}$ is 50\% higher than the COSMOS results. Some possible reasons
for the discrepancy in the value of f$_{SB}$ may be the different use
of inclination cuts (they used $i >$60$^o$ compared to our cut at $i
>$ 60$^o$) and selection criteria.  We chose our sample based on a
galaxy luminosity with an evolving luminosity function, spectral
type/color and visual classification, whereas \citet{jogee04} chose
their sample based on a fixed (lower) luminosity, range of U-V colors
(which is similar to our cut in T$_{phot}$), and a Sersic parameter
from a single fit to the galaxies or a concentration index.  Given the
various uncertainties at hand, we conclude that the data presented by
\citet{jogee04} can be interpreted as being consistent with the more
significant decline seen in our sample.

\citet{elm04} also reported a constant bar fraction to $z\sim 1.1$
based on an analysis of 186 background galaxies larger than 10 pixels
in diameter in the multi-color ACS image of the Tadpole galaxy.  These
data points are shown with the the purple triangles in Figure
\ref{comparestuds}.  The data show a declining bar fraction from
$\sim30$\% to $\sim15$\% out to $z=0.8$ with $3-\sigma$ uncertainty,
and a rise in the bar fraction from $z=0.8-1.1$, which is beyond the
redshift investigated here\footnote{Four galaxies in their Fig. 10 at
  $z \sim1.8$ are incorrect because of a photometric redshift error -
  the corrected data point is shown here}.  Their conclusion that the
bar fraction is flat on average followed primarily from the second
rise at $z\sim1$; otherwise their fractions agree with ours to within
statistical uncertainties. Our results are also in line with a recent
analysis of the Hubble Ultra Deep Field where the bar fraction, shown
with the blue triangle, is $\sim$10\% \citep{elm05b} at z$\sim$1,
consistent with the previous HDF studies and the values obtained in
this paper.



Although we have attempted to put all the data from various studies
into context, we emphasize that it is not straightforward to make
direct comparisons because of different selection criteria and bar
identification methods between these studies.  These may be
responsible for some of the observed differences.  The main point to
note is that in nearly every study, the data have shown a decline in
the bar fraction, although the interpretations of the data have ranged
from a constant bar fraction to a dramatic paucity of bars at
z$\sim$1.  It is only with the COSMOS data set that we are able to
robustly quantify the decline in the bar fraction and show that the
evolution is a strong function of the galaxy luminosity, mass, color
and bulge dominance.


\subsection{Formation of the Hubble Sequence: Assembling the Spiral Galaxies}\label{dissol}

The declining bar fraction reported in this paper shows that at a
lookback time of 7 Gyr ($z=0.835$) only about one-fifth of L$^*$
spiral galaxies were barred, which is about one-third the present day
value. During the following 3 Gyr (from $z\sim0.8$ to $z\sim0.3$) the
bar fraction increased to roughly its present value. Only small
changes occurred in the last 4 Gyr (z$<$0.3).

This evolution can be understood within the framework of classical bar
formation theory. $N$-body simulations have long suggested that bars
form spontaneously in galactic disks, usually on relatively fast
dynamical timescales. There are, however, two ways of slowing this
down. One is to increase the halo mass fraction within the disk
radius, and the other is to heat up the disk \citet{athasell86}.
Self-consistent three-dimensional (3D) simulations essentially agree
with this, although the role of the halo is now understood to be more
complex, so that the final bar can be considerably stronger in cases
where it grows slower \citet{ath02,ath03a}.  Thus, the time it takes
for an unbarred disk galaxy to become barred can vary widely. In cold,
disk-dominated cases, the bar forms within a Gyr or less, but
sufficiently hot disks embedded in very massive halos can stay
unbarred several Gyrs. Such a delay might well explain the time
evolution of the barred galaxy fraction shown in Fig. 1. Furthermore,
observations show that the halo-to-disk mass ratio is higher in low
mass, low luminosity galaxies than in bright, massive galaxies
\citep{bosma04,kranz03} so that bars are expected to grow later in the
former, as we indeed find here.  Hence if galactic disks are formed
with a variety of velocity dispersions and a variety of halo-to-disk
mass ratios, there should be a continuous increase with time in the
barred fraction, as observed here. The slope of this evolution will
depend on the distribution of the initial disk and halo parameters.
On the contrary, if all galaxy disks were, in the relevant
time-period, similar, then the fraction of disk galaxies that are
barred would be more or less constant with time, or show only a very
small increase.

The preceding paragraphs assumed the existence of the appropriate set
of physical conditions needed to form bars from the classical disk
instability. Alternatively, bar formation may coincide with inner disk
growth. This, however, would imply considerable growth of the inner
disk even for z$\sim$0.84, which seems inconsistent with recent
results that show no evolution in the disk scale-lengths at z$<$1
\citep{ravindranath04,barden05,sargent07,sheth07b}.  A further point
to take into consideration is that bars may dissolve when a gaseous
component is included in the angular momentum exchange cycle, and/or
in the presence of a sufficiently massive, centrally concentrated
object, as e.g., a black hole
\citep{friedli93,berentzen98,fukuda00,bournaud02, shen04, athlam05,
  bournaud05}. Observational evidence for bar dissolution, however, is
at present rather sparse \citep{das03, block02}, while the amount of
mass necessary seems to be larger than what is currently observed for
central mass concentrations \citep{shen04,athlam05}. Nevertheless, it
is by no means clear that this mechanism is unimportant.

We have so far discussed only isolated galaxies. Let us now turn to
the effect of interactions and mergings. The number of interactions
are known to increase dramatically with redshift (e.g.,
\citealt{kartaltepe07} and references therein).  Interactions and
merging activity are most likely to influence (heat up) the less
massive galaxies.  It is precisely in such galaxies that we see
significantly lower bar fractions compared to the high mass galaxies
at the highest redshifts (Figure \ref{Barfracmass},
\ref{Barfracvshubble}).  Although indirect, there is observational
evidence that later type and less massive systems are dynamically
hotter.  The top row of panels in Figure 1 of \citet{kassin07} clearly
shows that late-type spirals and irregulars have larger disordered
motions compared to early-type spirals particularly at high redshifts.
These are precisely the type of systems within which we find fewer
bars.  Moreover, in the same figure, the higher mass galaxies also
have a higher fraction of ordered motions than disordered motion,
although the trend is hard to see in the relatively modest sample size
in the high redshift bins.  These data suggest that the lack of bars
may therefore be related to the dynamic hotness and the mass surface
density of these disks.  We are currently identifying bars and
measuring the bar fraction in this sample of galaxies and should be
able to provide a direct answer for the said hypothesis (Sheth et al.,
in preparation).

Simulations show that interactions speed up bar formation in direct
encounters, but have little effect in retrograde ones
\citep{toomre72,noguchi87,gerin90, steinmetz02}, in good agreement
with observations \citep{kormendy79, elmelm82}. Thus one might have
expected higher rates of bar formation at high $z$, where interactions
are common.  On the contrary, it is possible for mergings to destroy
or severely weaken the bar, without destroying the disk
(e.g. \citealt{berentzen03}, and references therein).  More modeling
needs to be done before we can say with any certainty what the {\sl
  combined} effect of interactions and mergings is. Note that we
discarded from our statistics obviously interacting systems based on
tidal features or obvious distortions. However, if a galaxy is weakly
interacting, it would be difficult to distinguish it from a
non-interacting system; this is already the case even in the local
Universe. So our sample of galaxies is most likely probing quiescent,
post-merger or weakly interacting disks.

\subsection{The Downsizing Signature in Formation of Galactic Structure}\label{downsizing}

Galaxy ``downsizing'' was coined by \citet{cowie96} to refer to an
evolutionary history in which the most massive galaxies formed first.
There is strong observational evidence for the downsizing phenomena:
the presence of massive systems at high redshifts (e.g.,
\citealt{daddi04,daddi05,scarlata06,scarlata07}), an order of
magnitude decline in the typical star formation rate
\citep{arnouts05}, a change in the star formation activity to lower
mass systems with decreasing redshifts
\citep{treu05,glazebrook04,fontana04,maier05,maier06,bundy06}, and the
decrease in characteristic luminosity of active galactic nuclei at low
redshifts \citep{pei95,ueda03}.

The results presented here show a downsizing signature in the
formation of bars.  The most massive, luminous and red galaxies have a
higher bar fraction in the highest redshift bin with the most massive
and luminous systems having a bar fraction close to the present-day
value.  The early presence of bars in these galaxies in the context of
bar formation (\S \ref{dissol}) suggests that these systems
``matured'' early, i.e. they became dynamically cool and sufficiently
massive to host bars at z$>$ 0.8.  In contrast, the lower mass systems
which are also bluer, acquired a majority of their bars at z$\sim$0.8.
Thus the downsizing phenomenon must be considered to be an effect more
fundamental than one concerned solely with the regulation of ongoing
star formation; it seems to be intimately connected with the dynamical
maturity of the stellar disk.

\subsection{The Co-Evolution of Bulges \& Bars}\label{bulgeevol}

Figure \ref{fbarvsbulge} shows that in our highest redshift bin there
is a somewhat higher fraction of bars in galaxies with more massive
bulges.  This is consistent with the structural downsizing discussed
in the previous section because galaxies with bulges are denser and
more evolved in the center than galaxies without bulges.  Bars and
bulges apparently form at about the same time, with later times for
lower mass galaxies. Some care is necessary in the interpretation of
this result, however, since bulges are an inhomogeneous class of
objects. In this paper, we defined the bulge as the component in the
central part that contributes extra light above an extrapolated
exponential fit to the outer part. This definition includes both
classical (3D) bulges and disk-like pseudo-bulges
\citep{kormendy04,ath05}. Pseudo-bulges form by gas inflow and star
formation. Because bars drive inflow \citep{sheth05, sakamoto99},
there should be a correlation between disky bulges and bars,
independent of dynamical downsizing. Bars may also contribute to the
growth of nuclear black holes if they correlate with bulges, because
there is a tight correlation between black hole mass, stellar velocity
dispersion, and luminosity of bulges
\citep{kormendy95,magorrian98,ferrarese00}.  It will be interesting to
test such differences with subsequent analysis of the bulges and bars
in the COSMOS data.



\section{Conclusions}

Bars are an important signpost of galaxy evolution because once a
galaxy disk is sufficiently massive, dynamically cold and rotationally
supported it forms a bar.  Therefore the evolution of the bar fraction
over time is an important indicator of the evolutionary history of
disk galaxies and the assembly of the Hubble sequence.  Using a
detailed analysis of 2,157 L$^*$ face-on, spiral galaxies from
0.0$<$z$<$0.84 in the COSMOS 2-square degree survey we have
investigated the evolution of the bar fraction over the last 7 Gyr.
We have undertaken an extensive and careful analysis of selection
effects (k-correction, surface brightness dimming, inclination,
spatial resolution, etc.)  which is detailed in the Appendix.  Our
main results are as follows:

$\bullet$ The bar fraction for L$^*$ galaxies drops from about 65\% in
the local Universe to about 20\% at z=0.84. Over this redshift range
the fraction of {\em strong} bars (SB) drops from about 30\% to under
10\%.  Thus at a lookback time of 7 Gyr, when the Universe was half
its present age, fundamental aspects of Hubble's `tuning fork'
classification sequence had not yet fallen into place.  Only about one
fifth of all spiral galaxies were ``mature'' enough (dynamically cold,
massive and rotationally supported) to host galactic structures of the
type we see today.

$\bullet$ For the total f$_{bar}$ (SB+SAB), the change is far less
dramatic between z=0.3 and z=0.0 indicating slow evolution in galactic
structures in L$^*$ galaxies over the last 4 Gyr.  It is likely that
there is significant evolution in the formation of bars in the
sub-L$^*$ galaxies over this period.

$\bullet$ One of the most significant findings in this study is the
correlation between f$_{bar}$ and the galaxy mass, luminosity and
color. We find that in the highest redshift bins f$_{bar}$ is higher
in the more massive, luminous and redder systems. In fact, in the most
massive systems, f$_{bar}$ is already as high at z=0.8 as the local
value.  These systems thus had already arrived with their present
Hubble types at a lookback time of 7 Gyr.  In the subsequent 3 Gyr,
from z=0.84 to z=0.3, the lower mass, bluer systems evolved more
slowly toward their present Hubble types.  Thus the signature of
downsizing is intimately connected with dynamical maturity of disks
and is present in the formation of galactic structure.

$\bullet$ Finally, we find a slight preference for barred galaxies to
be more bulge-dominated in the high redshift bin.  This correlation is
consistent with the dynamical downsizing found for bars in general if
bars and bulges both form earlier and more prominently in the most
massive galaxies. The lack of a stronger correlation may be related to
the variety of bulges: bars are also likely to be involved with the
inflow that builds pseudo-bulges.  Given the strong correlation
between bulge properties and black hole mass seen today, there may be
a co-evolution of bars, bulges and black holes in some galaxies.  The
exact details of these processes remain to be investigated.

\acknowledgments

We are indebted to the anonymous referee for many helpful comments and
suggestions that have greatly improved this paper.  We are also
thankful for the insightful and helpful discussions we have had with
Karin Menendez-Delmestre, Neal Evans, Lee Armus, Isaac Shlosman, Wyn
Evans, Donald Lynden-Bell, Mark Dickinson, David Elbaz, Francois
Schweizer, Tomasso Treu, Jason Melbourne, Dan Kelson and Luis Ho.

The HST COSMOS Treasury program was supported through NASA grant
HST-GO- 09822. We wish to thank Tony Roman, Denise Taylor, and David
Soderblom for their assistance in planning and scheduling of the
extensive COSMOS observations. We gratefully acknowledge the
contributions of the entire COSMOS collaboration consisting of more
than 70 scientists. More information on the COSMOS survey is available
at http://www.astro.caltech.edu/$\sim$cosmos. It is a pleasure the
acknowledge the excellent services provided by the NASA IPAC/IRSA
staff (Anastasia Laity, Anastasia Alexov, Bruce Berriman and John
Good) in providing online archive and server capabilities for the
COSMOS datasets. We also wish to acknowledge support for the COSMOS
Science meeting in May 2005 which was supported in part by NSF grant
OISE-0456439.

{\sl Facilities}: HST (ACS).

\bigskip
\appendix
\bigskip
\centerline{\Large\bf Appendices}

\section{ANALYSIS OF SELECTION EFFECTS}
\label{seleffects}

Although we have carefully chosen a robust sample of galaxies, used
multiple methods for identifying bars and analyzed a sample of local
SDSS galaxies in the same manner as the COSMOS galaxies (\S
\ref{selection}), our results are in contradiction to some previous
studies.  Therefore we do additional investigation of the remaining
possible selection effects (cosmological / surface brightness dimming
and spatial resolution), which might produce a declining bar fraction.

\subsection{K-Corrections and the Effects of Bandshifting}\label{bandshifting}

The ACS data for COSMOS utilizes the broad F814W filter which traces
different rest-frame wavelengths at different redshifts.  As a result
it is imperative to understand the effects of k-correction
(bandshifting) and correct them as necessary.  To quantify the effects
of k-correction on the identification of bars, we examined a local
sample of 139 galaxies in all five Sloan bands ({\sl u,g,r,i and z}).  We
selected nearby ($<$ 100 Mpc), face-on (b/a $>$ 0.58), large (a 90\%
radius $>$ 2 kpc) and bright (M$_B  $ -19.7, M$_B$ estimated from
$g-r$ colors and $g$-band magnitudes \citealt{blanton03}) spiral
galaxies from the Sloan Digital Sky Survey (SDSS)
\citep{york00,gunn98} Data Release 4 \citep{adelman06}.  The data were
mosaicked and calibrated using the methods described by
\citet{west07}.

\begin{figure*}
\epsscale{1.}
\plotone{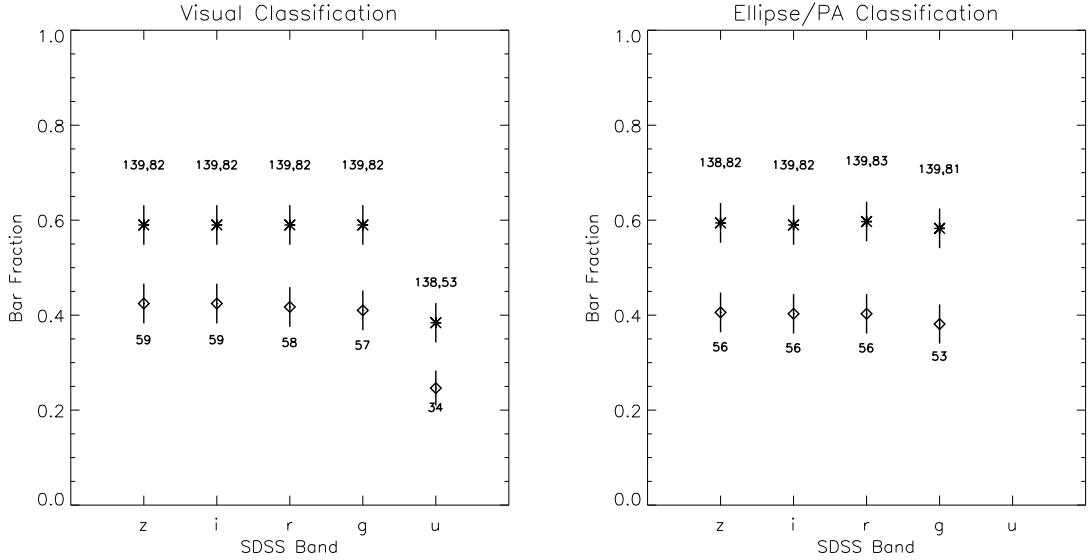}
\caption{Bar fraction as a function of the SDSS filters (u,g,r,i and
  z) for a local sample of 139 SDSS galaxies.  The left panel shows
  the results using the visual classification method and the right
  panel using the ellipse fitting method described in \S
  \ref{methods}.  These are the same methods used for the analysis of
  the COSMOS data.  The asterisks show the total bar fraction and the
  diamonds show the strong bar fraction, as described in \S
  \ref{methods}.  The pair of numbers above each point are the total
  number of galaxies and bars identified by each method.  The u band
  data point is missing in the right panel because the ellipse
  fitting algorithm fails in a majority of the galaxies in the u band.
  As noted in the text the main point of this exercise is to quantify
  the effects of k-correction on bar identification.  We find that
  there is a significant k-correction for the bar fraction short wards
  of the Balmer break in the u-band but the bar fraction is constant
  from the z-band to the g-band.
  \label{sdssbarfrac}}
\end{figure*}

We used the same bar identification methods (ellipse fitting \& visual
classification, \S \ref{methods}) for the SDSS data as for the COSMOS
data and measured the bar fraction in each band.  The results are
shown in Figure \ref{sdssbarfrac}.  The bar fraction is {\sl
  unchanged} from the z-band to the g-band at f$_{bar}\sim$0.6. This
is consistent with a number of previous studies (e.g.,
\citealt{menendez07, eskridge02, whyte02}) that have shown that the
overall bar fraction does not change appreciably between the optical
and near-infrared bands. This figure also demonstrates that the
shifting rest-wavelength of observation in our sample does not bias
our measurements of the bar fraction, provided we restrict the maximum
redshift of our sample appropriately.

\begin{figure}
\epsscale{1.}
\plotone{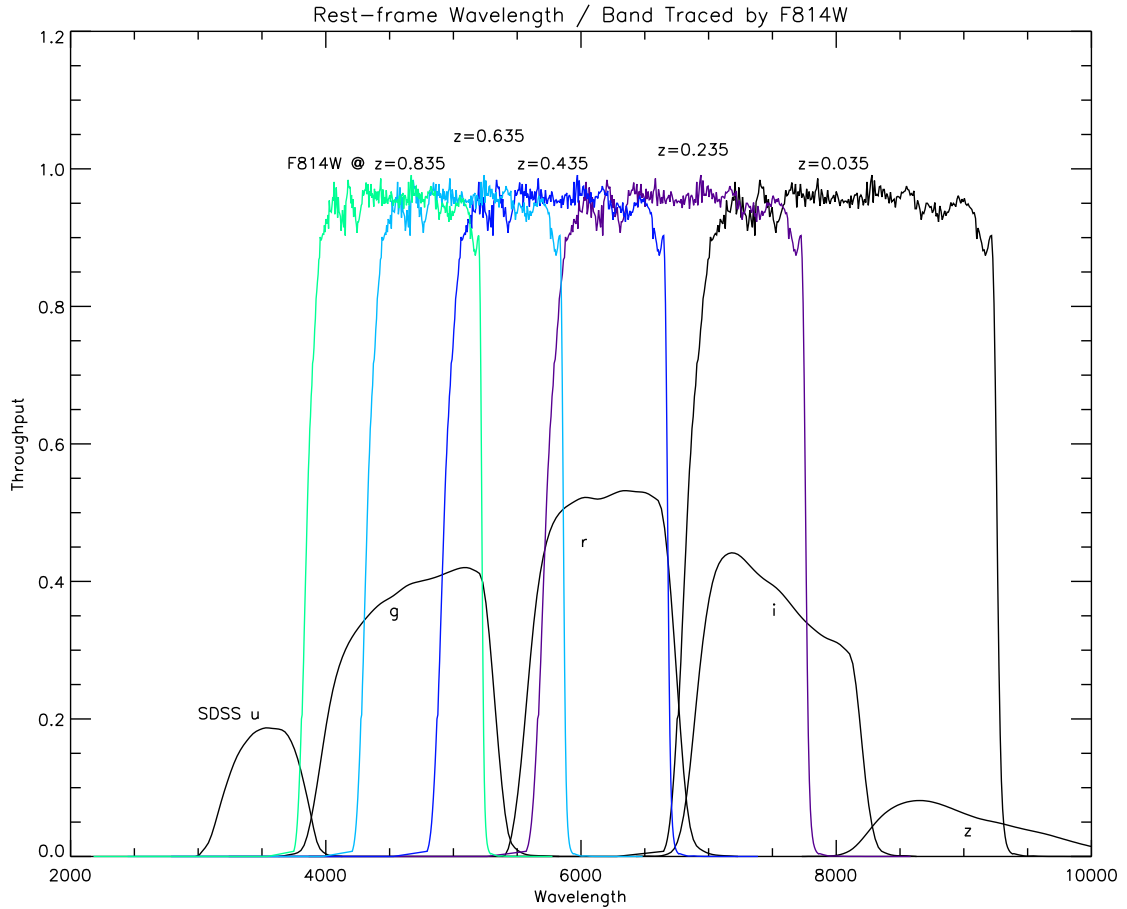}
\caption{The rest-frame wavelengths traced by the F814W filter over
  the redshift range of this study.  At the highest redshift z=0.835,
  the F814W filter is tracing the rest-frame SDSS g band filter where
  the effects of k-correction and identification of bars is still
  minimally affected as discussed in \S
  \ref{bandshifting}. \label{sdssfilters}}
\end{figure}

The maximum redshift chosen is important, because Figure
\ref{sdssbarfrac} shows that the SDSS bar fraction does appear to
decline markedly in the u-band. At this wavelength, in a majority of
cases, the ellipse fitting technique fails completely.  This is not
unexpected, and a component of this decline may find its origin in the
relatively poor signal-to-noise of the SDSS u-band data. However, we
suspect that the bulk of this decline is real.  Bars are primarily
stellar structures and some become significantly fainter and sometimes
disappear altogether short wards of the Balmer break.  A dramatic
example of this is shown for the nearby strongly barred spiral NGC
4303 in Figure 1 of \citet{sheth03}.  This is further justification
for our chosen limiting redshift in this paper, because by restricting
our sample to $z = 0.835$, the F814W filter does not probe bluer than
rest-frame g-band, as shown in Figure \ref{sdssfilters}.

\subsection{Objects with Peculiar Morphology}
\label{pec}

\begin{figure*}
\epsscale{0.85}
\plotone{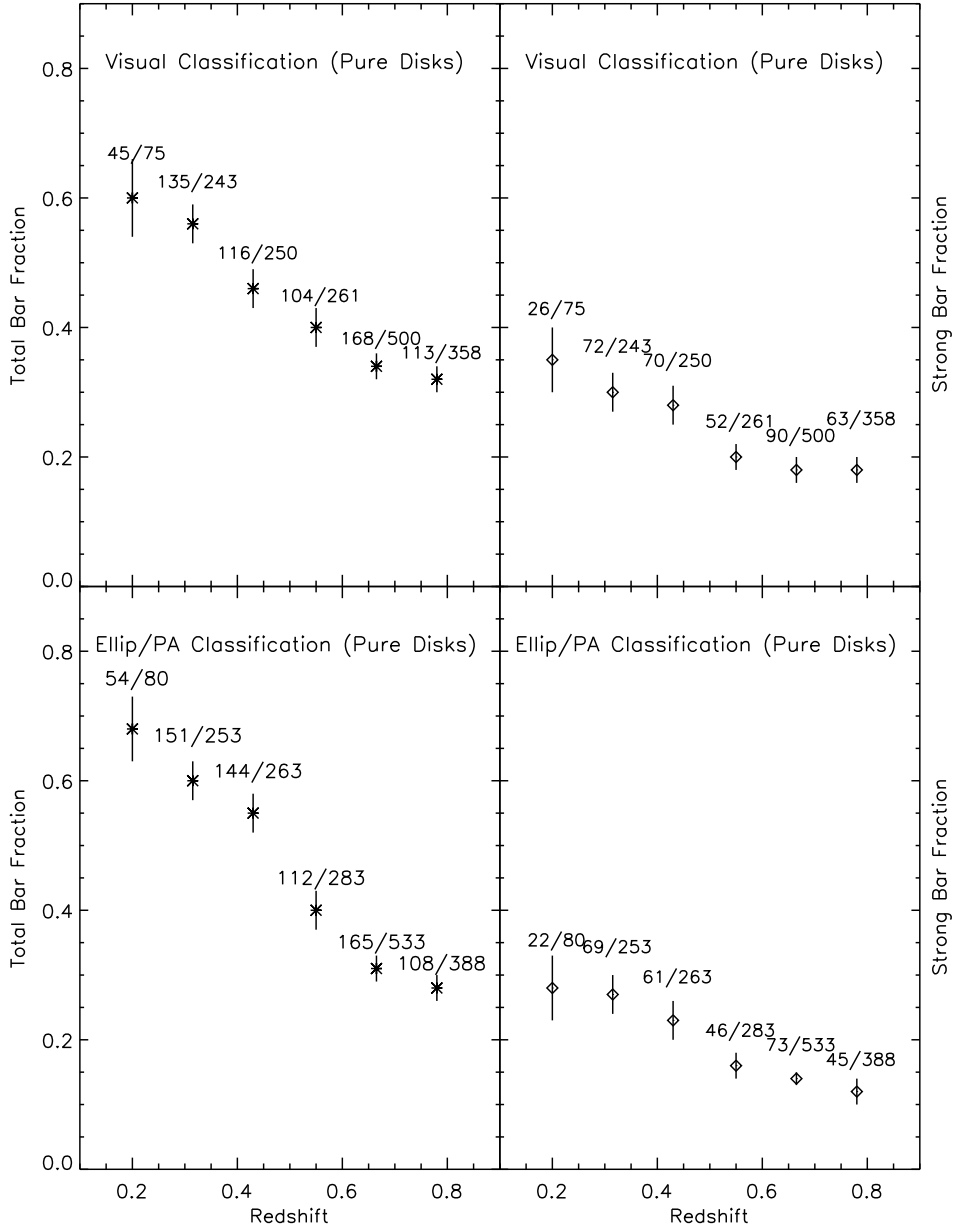}
\caption{As for Figure \ref{barfrac} except that galaxies with clump
  cluster, compact or other non-standard morphologies are discarded
  (\S3).  The overall trends noted in the discussion of Figure
  \ref{barfrac} remain the same.  \label{noudfbarfrac}}
\end{figure*}

We considered the possibility that the bar fraction may be incorrectly
measured at high redshifts due to the presence of a more exotic
variety of morphologies that have been observed at high redshifts.
\citet{elm07,elm06,elm05a,elm05b,elm05c} have identified
clump-cluster, compact, and chain galaxies at high redshifts which do not
have counterparts in the local Universe.  Could the f$_{bar}$ be
artificially lower because we are including more of these objects in
the total sample at the higher redshifts?  As noted earlier, this is
not the case for the visual classification method for which these
objects were already excluded.  We specifically exclude these objects
for the ElPa classification method and recompute the evolution of the
bar fraction.  The results are shown in Figure \ref{noudfbarfrac}.
The drop in the bar fraction in this revised sample is less steep as
expected.  f$_{bar}$ drops to 0.28$\pm$0.2 at z=0.835, and f$_{SB}$
drops to 0.12$\pm$0.02.  Both methods of classifying bars thus show
declines of 50\% in the bar fraction from the present day to z=0.835.

\subsection{Cosmological Surface Brightness Dimming}

Surface brightness dimming is critical even at z$<$ 1 because it
evolves so strongly with redshift ($\propto(1+z)^{4}$).  The
measurement of the bar fraction may be affected if the data are too
shallow because as the disk of the galaxy fades the bar, which usually
has a higher mean surface brightness, may remain visible and therefore
be misclassified as an inclined spiral (e.g., \citealt{jogee02}).  To
investigate this possibility we investigated the ability of the COSMOS
data (and a few other surveys) to trace an outer disk (several times
the typical bar semi-major axis) isophote as a function of redshift.

The median bar semi-major axis measured from the 2MASS Large Galaxy
Atlas survey is a$_{bar}$=4.2$\pm$2.9 kpc and relative size is
a$_{bar}$ / R$_{25}$ = 0.29$\pm$0.17 \citep{menendez07}, where
R$_{25}$ is the classic 25 magnitude arcsec$^{-2}$ isophote in the
B-band.  Thus the R$_{25}$ radius is at least {\sl three} times larger
than a typical bar.  In the outer regions of a galaxy, B-I = 1.5.  So
the corresponding limiting I-band isophote is at $\mu_I$ = 23.5.  An
isophote a full magnitude fainter ($\mu_I$=24.5 magnitudes
arcsec$^{-2}$) can therefore be safely considered to be an outer disk
isophote.

\begin{figure*}
\epsscale{1.}
\plotone{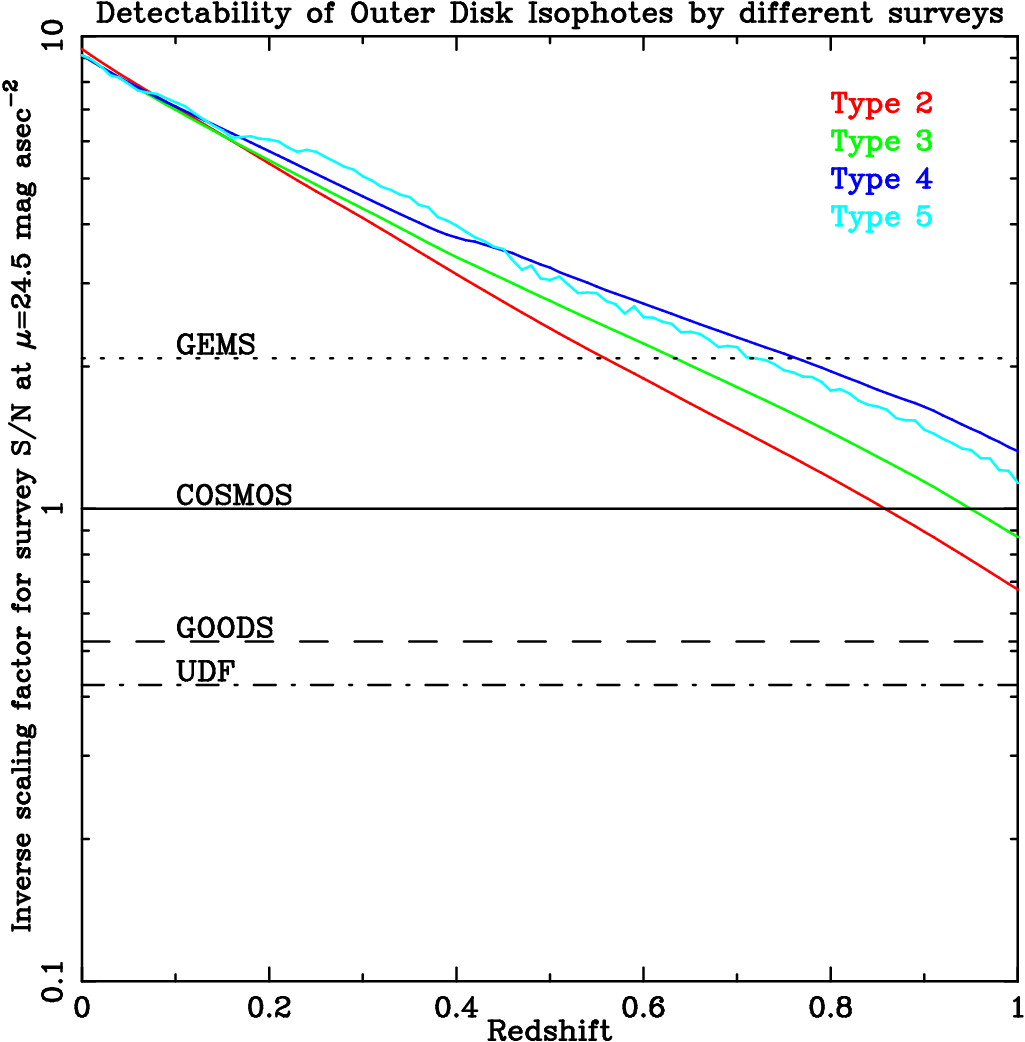}
\caption{A plot of the noise to signal ratio for 0.6\arcsec diameter
  galaxy with a rest-frame $\mu_{I_AB}$ = 24.5 magnitudes
  arcsec$^{-2}$ in COSMOS.  The choice of 0.6$\arcsec$ diameter was
  based on previous studies of galaxy sizes \citep{ferguson04}.  Our
  own data are consistent with these size estimates.  We find that the
  median exponential scale length of L$^*$ galaxies in COSMOS is 3.1
  kpc (0$\farcs$ 39) at z=0.835 \citep{sheth07b}.  In the sample
  analyzed here, 93\% of the galaxies have a half light radius greater
  than than 0.3$\arcsec$.  So the noise to signal measurement shown
  here for COSMOS data for the outermost disk isophote is a lower
  limit.  The horizontal lines show the $1\sigma$ sensitivity limits
  for the COSMOS, GEMS, GOODS, and UDF survey.  Note that GEMS will be
  adversely affected in measurements f$_{bar}$ even for L$^*$ galaxies
  at $z > 0.5$ because of the low S/N in the underlying disk.
  \label{surfbright1}}
\end{figure*}

In Figure \ref{surfbright1} we show the noise to signal value reached
for a $\mu_I$=24.5 magnitudes arcsec$^{-2}$ isophote fading due to
cosmological surface brightness dimming.  The differently colored
lines are for the different SED types.  It is clear from this figure
that COSMOS, GOODS, HDF and UDF are sufficiently deep to allow one to
detect the outer edges of a typical L$^{*}$ disk to z$\sim$1.  This is
not the case for the shallower GEMS data which are unable to quantify
the outer disk isophotes for L$^*$ galaxies at z $>$ 0.5.

In this calculation we assume that the disks have evolved passively
from z=1 to the present, when in fact there is a magnitude of
luminosity evolution in disks from the increased star formation rate.
Therefore choosing a luminosity evolution in the sample selection
criteria, as we have done for the COSMOS data, further improves the
signal to noise and ability to confidently trace the outermost disk
isophotes with redshift.  Based on this analysis we are confident that
disk dimming is not responsible for the observed decline in the bar
fraction with redshift in the COSMOS data.

\begin{figure*}
\epsscale{1.1}
\plotone{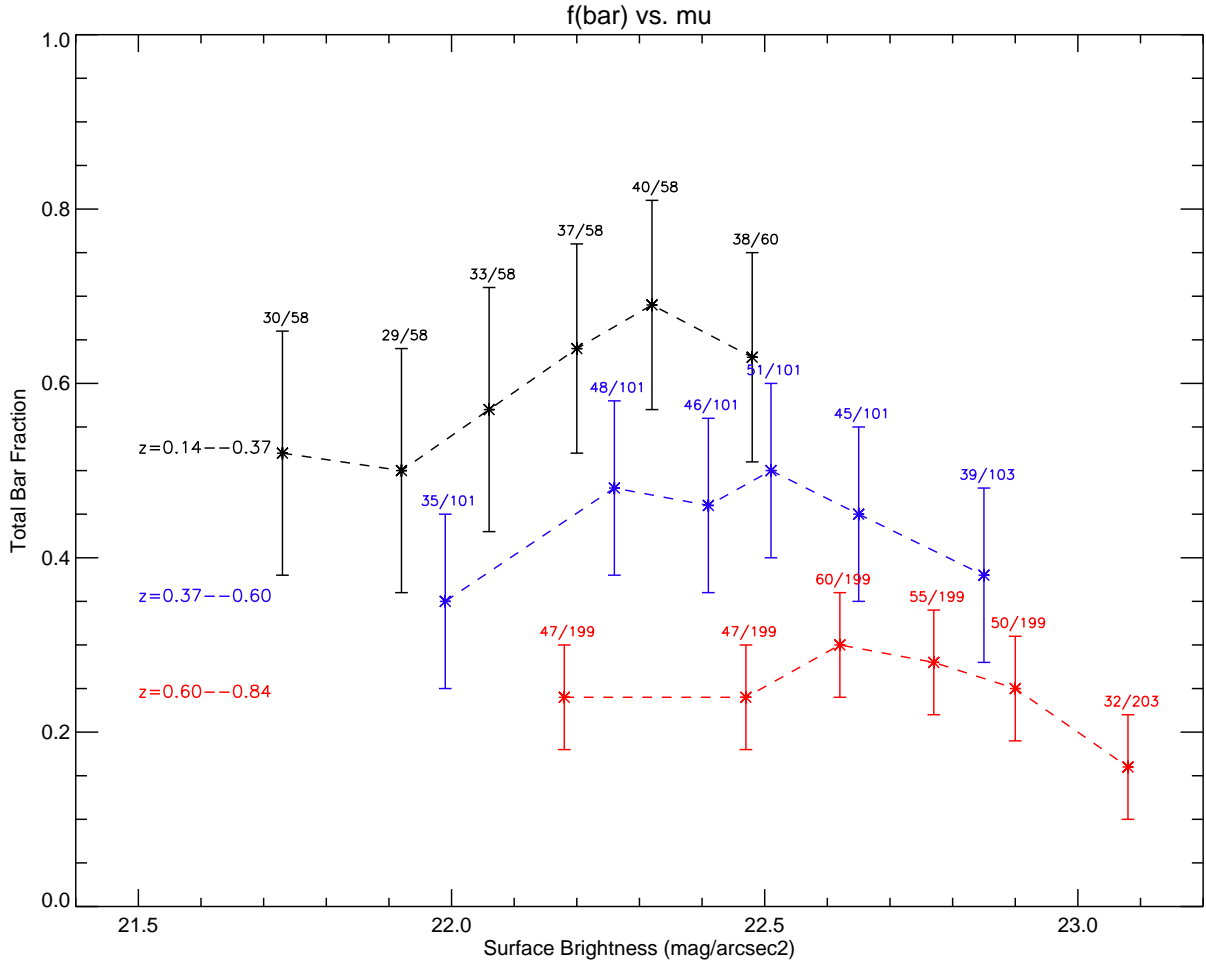}
\caption{A plot of the total bar fraction (f$_{bar}$) versus measured
  surface brightness in the COSMOS sample in three redshift bins.  We
  find that there is no correlation between the bar fraction and the
  surface brightness of the galaxies, as might have been expected if
  surface brightness dimming was affecting the measurement of the bar
  fraction.  The overall trend of lower bar fraction with redshift can
  be seen readily in these panels.  The results are the same when
  considering only the strong bar fraction.
 \label{surfbright2}}
\end{figure*}

We further tested the effects of surface brightness dimming on the
data with a second empirical check by comparing the bar fraction as a
function of the observed surface brightness of the disks in each
redshift bin.  This test is shown in Figure \ref{surfbright2}.  If
indeed surface brightness dimming was responsible for a decline in the
bar fraction and we were preferentially classifying lower surface
brightness disks as unbarred, we should see a correlation of f$_{bar}$
with $\mu$.  We see no trends in the bar fraction with $\mu$ in any
redshift bin.  Thus we conclude that the observed evolution of the bar
fraction shown in Figure \ref{barfrac} is not due to cosmological
surface brightness dimming of the disks.


\subsection{Spatial Resolution and the Bar Fraction}

One possibility for the observed decline in the bar fraction could be
that we are preferentially missing small bars at higher redshifts.  As
already noted earlier, (and as shown in Figure 3 of \citealt{sheth03})
ACS resolution (1 pixel = 0$\farcs$05 = 0.38 kpc at z=0.835) is
adequate for identifying all bars larger than 2 kpc at all
z$<$0.835. Bars smaller than this, at least in the local Universe, are
nuclear bars and not the primary bars that we are concerned with in
this paper.  Very small bars in late type galaxies have been measured
by \citet{erwin05} in a local sample but it is unclear whether these
galaxies are comparable to the large, L$^*$ and brighter galaxies
being examined in our sample.  For L$^*$ galaxies, there is
substantial evidence that the size of galaxy disks does not change
significantly to z$\sim$1.  For the COSMOS sample, the median disk
scale length is unchanged over the redshift range under investigation,
consistent with similar results found previously
\citep{ravindranath04, barden05, sargent07}.  Given that the bar
semi-major axis is typically 0.3R$_{25}$, the lack of a change in disk
sizes also indicates that it is unlikely that bars change their sizes
significantly as a function of redshift.

\begin{figure*}
\epsscale{1.1}
\plotone{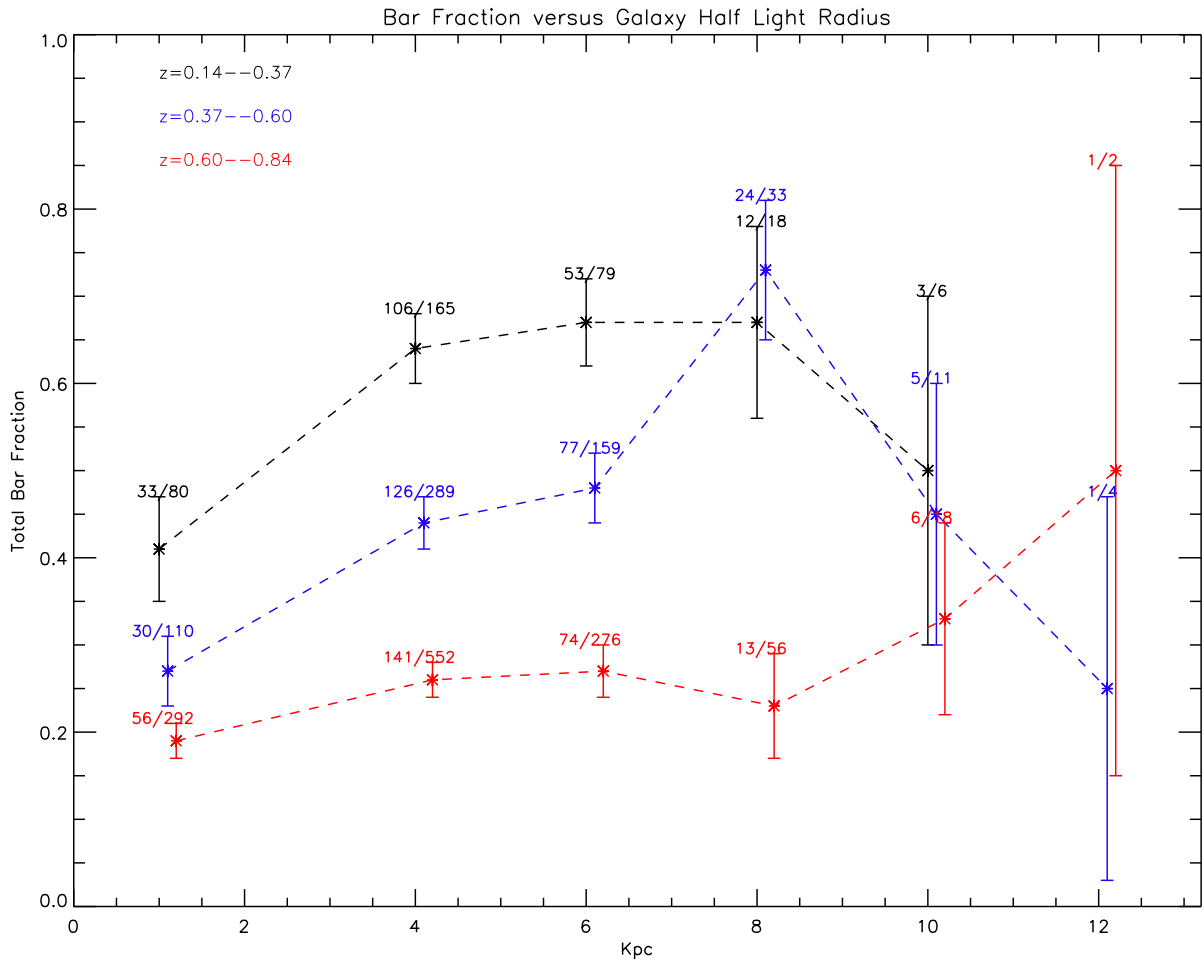}
\caption{A plot of the total bar fraction (f$_{bar}$) versus
  half-light radius.  If the bar fraction was indeed declining because
  of smaller galaxy disks with unresolved bars, one would expect to
  see a decreasing f$_{bar}$ with smaller disks particularly in the
  highest redshift bin.  We find no such trend and therefore conclude
  that the general decline in the bar fraction is not correlated with
  disk size.
 \label{Barfracvssize}}
\end{figure*}

As another check we decided to see if the measurement of the bar
fraction was affected by the size of galaxies in a given redshift bin.
If the bar fraction was indeed declining because of smaller galaxy
disks with unresolved bars, one would expect to see a decreasing
f$_{bar}$ with smaller disks and this effect would be most pronounced
at higher redshifts where the linear resolution of the ACS data is the
coarsest.  When we plot f$_{bar}$ versus the half light radius of
galaxies in different redshift bins in Figure \ref{Barfracvssize}, we
find no significant correlation between the bar fraction and the size
of the disk in any of the redshift bins.  There is a slight decline in
the bar fraction for the smallest galaxies.  This reflects the finding
that bars are less frequent in less massive systems and not a
selection effect.  The result is the same when considering only
f$_{SB}$, or using the exponential scale length for an estimate of the
galaxy disk size.  We therefore conclude that there is {\sl not} a
preferential loss of bars in disks of any particular size at any
redshift in this study.

\section{Artificially Redshifting Galaxies}\label{simul}

\begin{figure*}
\epsscale{0.9}
\plotone{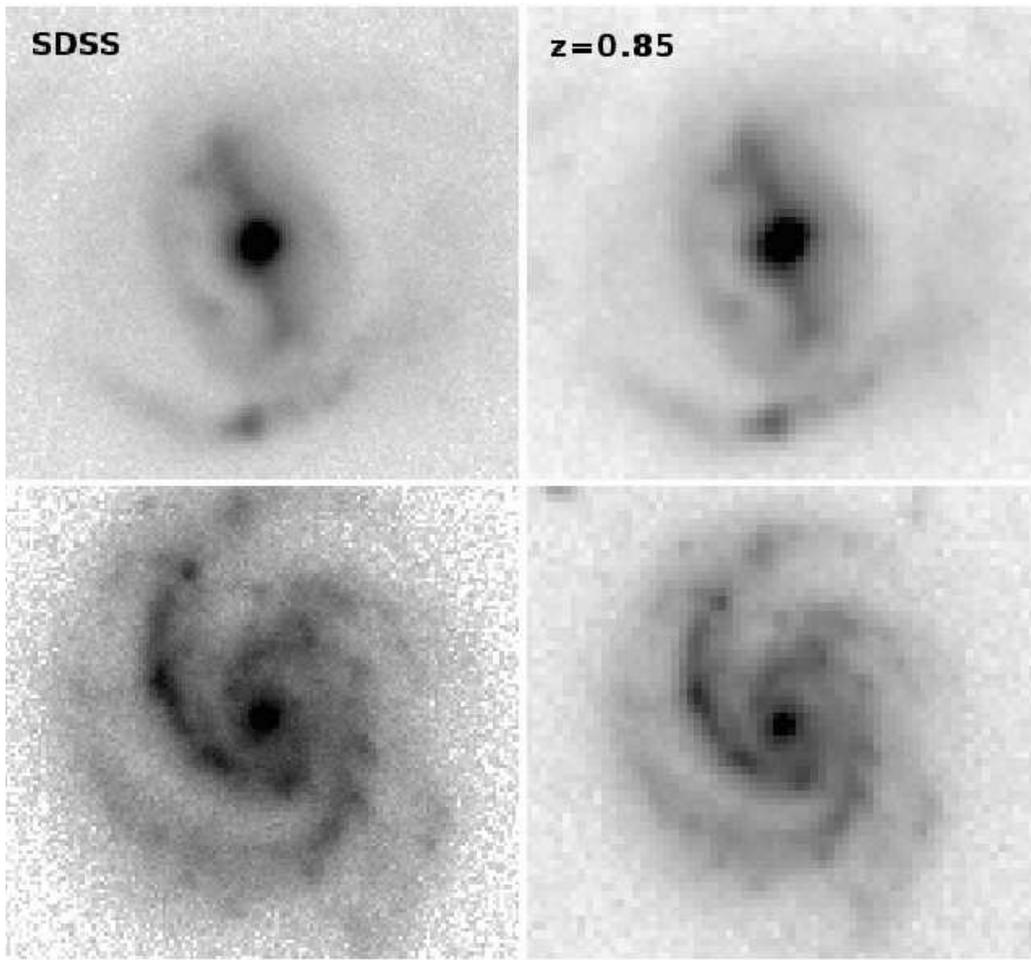}
\caption{Two examples of artificially redshifted SDSS galaxies.  A
  barred spiral is shown in the top row and a spiral galaxy in the
  bottom row.  The left column shows the original SDSS g-band image
  and the right column is the artificially redshifted image at z=0.84
 \label{artif}}
\end{figure*}
The above sections have addressed most of the classical selection
effects that plague high redshift studies, namely k-correction,
resolution, surface brightness dimming, etc.  To conclusively test all
of these effects we artificially redshifted the g-band images of all
139 SDSS galaxies to z=0.84 following the technique outlined in
\citet{giav96}.  We redshifted the galaxies to the F814W filter,
rebinned the image, took into account surface brightness dimming and
matched the noise characteristics of our ACS data.  Examples of the
artificially redshifted galaxies are shown in Figure \ref{artif}.  We
re-classified these images into strongly barred, weakly barred and
unbarred spirals as we had done before.  In 127/139 galaxies the
classification remained unchanged.  Of the remaining twelve galaxies,
seven are now classified as weakly barred in these images compared to
the original SDSS images where they were classified as unbarred.  Four
galaxies classified previously as weakly barred are now classified as
unbarred.  And one galaxy that was classified as strongly barred is
now classified as an unbarred spiral.  So overall the bar fraction did
not change appreciably (original SDSS images f$_{bar}$=0.59,
artificially redshifted SDSS images f$_{bar}$=0.60).  These results
are remarkable but not unexpected and reflect the exquisite
sensitivity of the ACS survey, which was designed to be as deep at z=1
as the SDSS survey is locally.

\clearpage
\begin{deluxetable}{rrrrrrrrrr}
\tablecolumns{10}
\tablewidth{0pc}
\tablecaption{Galaxy Classification \& Bar Fraction with Redshift \label{tab1}}
\tablehead{
\colhead{z$_L$\tablenotemark{a}} &\colhead{z$_U$\tablenotemark{b}} &\colhead{N$_{bin}$\tablenotemark{c}} &\colhead{SP$_{bin}$\tablenotemark{d}} &\colhead{WB$_{bin}$\tablenotemark{e}} &\colhead{SB$_{bin}$\tablenotemark{f}}  &\colhead{f$_{bar}$\tablenotemark{g}} &\colhead{1$\sigma$\tablenotemark{i}} &\colhead{f$_{SB}$\tablenotemark{h}} &\colhead{1$\sigma$\tablenotemark{i}}
}
\startdata
\multicolumn{10}{c}{ Visual Classification} \\
\tableline
  SDSS-i & 0.00 & 139 & 57 & 23 & 59 & 0.59 & 0.04 & 0.42 & 0.04 \\
  0.14 &  0.26 &  75 &  30 &  19 &  26 &  0.60 &  0.06 &  0.35 &  0.05 \\
  0.26 &  0.37 & 243 & 108 &  63 &  72 &  0.56 &  0.03 &  0.30 &  0.03 \\
  0.37 &  0.49 & 254 & 136 &  46 &  72 &  0.46 &  0.03 &  0.28 &  0.03 \\
  0.49 &  0.61 & 265 & 161 &  52 &  52 &  0.39 &  0.03 &  0.20 &  0.02 \\
  0.61 &  0.72 & 504 & 336 &  78 &  90 &  0.33 &  0.02 &  0.18 &  0.02 \\
  0.72 &  0.84 & 364 & 251 &  50 &  63 &  0.31 &  0.02 &  0.17 &  0.02 \\
\tableline
\multicolumn{10}{c}{Ellipse/PA Classification} \\
\tableline
 SDDS-i & 0.00 & 139 & 57 & 26 & 56 & 0.59 & 0.04 & 0.41 & 0.04 \\
 0.14 &  0.26 &  83 &  29 &  32 &  22 &  0.65 &  0.05 &  0.27 &  0.05 \\
 0.26 &  0.37 & 267 & 114 &  84 &  69 &  0.57 &  0.03 &  0.26 &  0.03 \\
 0.37 &  0.49 & 282 & 134 &  85 &  63 &  0.52 &  0.03 &  0.22 &  0.02 \\
 0.49 &  0.61 & 326 & 210 &  68 &  48 &  0.36 &  0.03 &  0.15 &  0.02 \\
 0.61 &  0.72 & 668 & 495 &  97 &  76 &  0.26 &  0.02 &  0.11 &  0.01 \\
 0.72 &  0.84 & 530 & 412 &  71 &  47 &  0.22 &  0.02 &  0.09 &  0.01 
\enddata
\tablenotetext{a}{~ z$_L$: Lower limit of redshift bin}
\tablenotetext{b}{~ z$_U$: Upper limit of redshift bin}
\tablenotetext{c}{~ N$_{bin}$: Total number of spiral galaxies in bin}
\tablenotetext{d}{~SP$_{bin}$: Number of galaxies classified as unbarred in bin}
\tablenotetext{e}{~ WB$_{bin}$: Number of galaxies classified as weak bars (SAB) in bin}
\tablenotetext{f}{~ SB$_{bin}$: Number of galaxies classified as strong bars (SB) in bin}
\tablenotetext{g}{~ f$_{bar}$ = WB$_{bin}$ $+$ SB$_{bin}$ / N$_{bin}$}
\tablenotetext{h}{~ f$_{SB}$ = SB$_{bin}$ / N$_{bin}$}
\tablenotetext{i}{~ Error assuming binomial statistics. 1$\sigma$=$\sqrt (f(1-f)/N_{bin}$), where f is f$_{bar}$ or f$_{SB}$}

\end{deluxetable}

\end{document}